\documentclass
[preprintnumbers,showkeys,showpacs,nofootinbib,byrevtex]{revtex4}
\usepackage{amsmath,amsfonts,amssymb,amscd,amsxtra,amsthm}
\usepackage{graphicx}
\usepackage{bm}
\begin{document}
\title{Charged pion photoproduction with the $\Delta(1232)$ baryon beyond the resonance region}
\author{Seung-il Nam}
\email[E-mail: ]{sinam@kau.ac.kr}
\affiliation{Research Institute for Basic Sciences, Korea Aerospace University, Goyang 412-791, Korea}
\author{Byung-Geel Yu}
\email[E-mail: ]{bgyu@kau.ac.kr}
\affiliation{Research Institute for Basic Sciences, Korea Aerospace University, Goyang 412-791, Korea}
\date{\today}
\begin{abstract}
We investigate the charged pion photoproduction off the proton
target with the $\Delta(1232)$ baryon in the final state, i.e.
$\gamma p\to\pi^{-}\Delta^{++}(1232)$ and $\gamma
p\to\pi^{+}\Delta^{0}(1232)$, based on the effective Lagrangian
method, beyond the resonance region, $E_\mathrm{cm}\gtrsim2$ GeV.
We employ the $\pi$- and $\rho$-meson Regge trajectories in the
$t$-channel, in addition to the proton- and $\Delta$-pole, in the
$s$- and $u$-channels respectively, and the contact-interaction
contributions. A specific scheme for the form factor which
satisfies the Ward-Takahashi identity, crossing symmetry, and
on-shell condition, is taken into account. To discuss the validity of the Regge approach within the Effective Lagrangian method, we also consider a smooth interpolation between the
Regge- and Feynman-propagators for the $t$-channel meson exchanges
as a function of $\sqrt{s}$. We present the numerical results for
the energy and angular dependences of the cross sections, and
double-polarization observable. It turns out that the present
framework  shows the significance of the $\pi$-exchange and
contact interaction terms to reproduce the experimental data
qualitatively well. Especially, the interpolation between the two
propagators plays a crucial role to reproduce the high-energy
experimental data. The $\pi^+$ decay-angle distribution is also
studied using the $\Delta^{++}$-decay frame, i.e. the
Gottfried-Jackson frame. The present results will be a useful
guide for future high-energy photon-beam experiments.
\end{abstract}

\maketitle
\section{Introduction}

Hadron productions via various scattering processes have been one
of the most important experimental and theoretical methods to
investigate the strongly-interacting systems, which are governed
by the fundamental theory, i.e. {\it quantum chromodynamics} (QCD), in
terms of the color-singlet degrees of freedom. Among the various
production processes, photo and electro-productions have been
proved to be very useful. Since the photon is a clear probe, these
production processes are well suited for investigating hadronic
properties, such as the structures, the reaction mechanisms, and
so on.  Moreover, the nonstrange-meson electro and
photoproductions are an extremely good method to investigate the
baryon-resonance search. On top of these specific features,
according to the gauge-boson nature of the photon, the photocoupling constrains the scattering processes in such a way that the
Ward-Takahashi (WT) identity must be satisfied to all orders of
Feynman diagrams. Due to this constraint, one can simply identify the necessary Feynman diagrams for a certain scattering process.

Experimentally and theoretically, meson electro and
photoproductions have been well studied in various ways. For
instance, in the previous work~\cite{Nam:2005uq}, employing the
Rarita-Schwinger formalism for spin-$3/2$ fermions, it turns out
that the contact term interaction, which is responsible for
conserving the WT identity, plays a dominant role for the
$\Lambda(1520)$ photoproduction. If this is the case, there
appears the difference of the production rate  between those from
the proton and neutron targets related to their isospin structures
in each reaction process. Interestingly enough, this theoretical
consequence has been confirmed by the experiments~\cite{Nakano},
and supported theoretically~\cite{Toki:2007ab}. We also note that
the formalism used in
Refs.~\cite{Nam:2005uq,Nam:2009cv,Nam:2010au} reproduced the
presently available data qualitatively very
well~\cite{Nam:2010au}. Moreover, the recent beam-energy upgrades
of the experimental facilities, such as the CLAS12 at Jefferson
laboratory~\cite{CLAS12} and the LEPS2 at SPring-8~\cite{LEPS2},
may shed light on the measurements for new high-energy data for
various photo and electroproduction processes.

In consideration with the success of the theoretical framework
employed in Refs.~\cite{Nam:2005uq,Nam:2009cv,Nam:2010au} and the
present experimental situation mentioned above, we are motivated to investigate
the charged pion ($\pi^{\pm}$) photoproduction off the proton target with $\Delta(1232,3/2^+)$ baryon,  $\gamma p\to\pi^{-}\Delta^{++}$ and $\gamma p\to\pi^{+}\Delta^{0}$,
beyond the resonance region $E_\mathrm{cm}\gtrsim2$ GeV. Although
there are other isospin channels  in the final $\pi\Delta$ state,
we would like to focus on these charged pion productions in the present
work, because of the abundant experimental data for these specific
channels. It is also worth mentioning that these elementary processes
are important ingredients to study two pion photoproduction
$\gamma N\to\pi\pi N$. As for the low-energy region, the photo
and electroproductions for these elementary-reaction process were
already investigated in Refs.~\cite{Nacher:1998hh,Ripani:2000va}.
In those works, it turns out that the
reaction process contains various contributions, such as those of
the Born terms including the nucleon and baryon resonances, the
final- (FSI) and initial-state (ISI) interactions, and the Regge
poles. However, as will be shown in the later Sections of the
present work, only the Born terms and Regge pole contributions
almost saturate the reaction process in the high-energy region
beyond the resonance region $E_\mathrm{cm}\gtrsim2$ GeV. In,
Ref.~\cite{Clark:1977ce}, the scattering amplitude for $\gamma
N\to\pi \Delta$ was also parameterized phenomenologically by a simple
one-pion exchange. The Regge poles and absorption corrections
were employed for the $\Delta$-photoproduction in Ref.~\cite{Goldstein:1974mr}

As a theoretical  framework, as done in
Refs.~\cite{Nam:2009cv,Nam:2010au}, we make use of the effective
Lagrangian method and mesonic Regge trajectories.  Since we are
interested in the energy range beyond the resonance region as
mentioned, we will not consider the resonant contributions in the
present work for simplicity. Thus, our strategy to investigate the
$\Delta$ photoproduction is quite simple: We take into account the
$\pi$- and $\rho$-meson Regge trajectories in the $t$ channel for
describing the high-energy experimental data, in addition to the
tree-level Born terms, such as the the proton- and $\Delta$-pole,
in the $s$- and $u$-channels respectively, and contact-term
contributions. In treating $\Delta$ theoretically, we make use of
the Rarita-Schwinger vector-spinor formalism~\cite{Rarita:1941mf}.
In addition to these ingredients, we also take into account an
interpolation between the Feynman and Regge propagators in the t-channel meson-exchange, considering that the Regge contributions can still affect on the process even in the low energy
region. We utilize an ansatz devised for this purpose, which was
proved essential to reproduce the low- and high-energy data
simultaneously~\cite{Nam:2010au}. All the relevant scattering
amplitudes are constructed in terms of the tree-level Born
approximation without resonant contributions as mentioned. To
consider the spatial  distributions of the hadrons involved, we
introduce the hadron form factors in a gauge-invariant scheme
which preserves the Ward-Takahashi identity with the Regge
propagators as done in~\cite{Nam:2010au,Ozaki:2009wp}.

From the numerical analyses, we will compute the energy and
angular dependences of the cross sections, polarization-transfer
coefficients, and $\pi^+$ decay-angle distribution of the process
under various conditions. It turns out that unpolarized physical
quantities, such as the energy and angular dependences are
reproduced qualitatively well beyond the resonance region
$E_\mathrm{cm}\gtrsim2$ GeV as compared to the presently available
experimental data. We also discuss the appearance of the strong
peak in the forward-scattering region which is decreasing as the
photon energy increases. It turns out that this forward peak is
generated by the $t$-channel pseudoscalar-meson ($\pi$) exchange.
The angular dependence $d\sigma/dt$ shows that the present
framework is not enough to reproduce the experimental data for the
region, where $E_\gamma\lesssim3.5$ GeV and
$-t\gtrsim0.05\,\mathrm{GeV}^2$. Nevertheless the present
theoretical framework provides a very good agreement with the
experimental data for the high-energy region beyond
$E_\gamma\approx5$ GeV. Especially, the interpolation of the
Feynman and Regge propagators plays an important role to reproduce
the data appropriately. From the numerical results for the $\pi^+$
decay-angle distribution, e find that the contribution of the $\rho$-meson exchange is small, but the contact-term contribution and the $\pi$-exchange are significant both of which compete with each other in the relatively low energy region. In
contrast, only the contact-term contribution dominates the process in the high-energy region.

The present work is  structured as follows: In Section II, we
briefly explain the theoretical formalism such as the effective
Lagrangians for the relevant interactions, gauge-invariant
scheme for the form factors, and mesonic Regge trajectories to compute
the reaction process which we are interested in. Numerical results
and related discussions are given in Section III. We summarize and
close with a conclusion in Section IV.

\section{Formalisms}
In this Section, we  explain the theoretical framework for the
present calculations for the $\gamma p\to \pi\Delta$ reaction
process. First, we present the relevant Feynman diagrams for the
present reaction process at the tree-level Born approximation in 
Fig.~\ref{FIG0}. The four momenta of
the particles involved are also given there. We will consider the
$(s,u,t)$-channel baryon and meson exchanges, and the contact-term
contributions. The contact term is necessary for preserving the WT
identity of the scattering amplitude as shown in
~\cite{Nam:2005uq}. The interaction Lagrangians for each vertex
are defined as follows:
\begin{eqnarray}
\label{eq:LAG}
{\cal L}_{\gamma PP}&=&
ie_{P}\left[(\partial^{\mu}P^{\dagger})P
-(\partial^{\mu}P)P^{\dagger}\right]A_{\mu},
\cr
\mathcal{L}_{\gamma P V}&=&
g_{\gamma PV}\epsilon_{\mu\nu\sigma\rho}(\partial^{\mu}A^{\nu})
(\partial^{\sigma}V^{\rho})P+\mathrm{h.c.},
\cr
\mathcal{L}_{\gamma NN}&=& -\bar{N}\left[e_N\rlap{\,/}{A}
-\frac{e\kappa_N}{4M_N}\sigma^{\mu\nu} F_{\mu\nu}\right]N, \cr
\mathcal{L}_{\gamma\Delta\Delta}&=&
-\bar{\Delta}^{\mu}
\left[\left(F_{1}\rlap{/}{\epsilon}g_{\mu\nu}
-F_3\rlap{/}{\epsilon}\frac{k_{1\mu}k_{1\nu}}{2M^{2}_{\Delta}}\right)
-\frac{\rlap{/}{k}_{1}\rlap{/}{\epsilon}}
{2M_{\Delta}}\left(F_{2}g_{\mu\nu}-F_4\frac{k_{1\mu}k_{1 \nu}}
{2M^{2}_{\Delta}}\right)\right]
\Delta^{\nu},
\cr
\mathcal{L}_{P N\Delta}&=&\frac{g_{PN\Delta}}{M_{P}}
\bar{\Delta}^{\mu}\partial_{\mu}P N+\mathrm{h.c.},
\cr \mathcal{L}_{VN\Delta}&=&
-\frac{ig^{(1)}_{VN\Delta}}{m_{V}}\bar{\Delta}^{\mu}\gamma^{\nu}\gamma_{5}
V_{\mu\nu}N
-\frac{g^{(2)}_{VN\Delta}}{m^{2}_{V}}\bar{\Delta}^{\mu}\gamma_{5}
V_{\mu\nu}\partial^{\nu}N
+\frac{g^{(3)}_{VN\Delta}}{m^{2}_{V}}\bar{\Delta}^{\mu}\partial^{\nu}
\gamma_{5}V_{\mu\nu}N+\mathrm{h.c.},
\cr
\mathcal{L}_{\gamma PN\Delta}
&=&-\frac{ie_Ng_{PN\Delta}}{M_{P}}
\bar{\Delta}^{\mu}
A_{\mu}PN+\mathrm{h.c.},
\end{eqnarray}
where $e_h$ stands for  the electric charge of the hadron $h$,
whereas $e$ is the unit electric charge. $A$, $P$, $V$, $ N$, and
$\Delta$ denote the fields for the photon, pseudoscalar and vector
mesons, nucleon, and $\Delta$, respectively. The magnetic moment
of the proton is $2.79\mu_N$, giving $\kappa_p=1.79$ for the
present case~\cite{Nakamura:2010zzi}. The antisymmetric tensor is
defied by
$\sigma=i(\gamma_{\mu}\gamma_{\nu}-\gamma_{\nu}\gamma_{\mu})/2$,
and the $F_{\mu\nu}$ and $V_{\mu\nu}$ are the field-strength
tensors for the photon and vector meson, respectively. Note that
$F_{1\sim4}$ are the multipole moments of the $\Delta$,
corresponding to the monopole, dipole, quadrupole, and octupole
ones. Since there are no experimental and theoretical information
for the $F_3$ and $F_4$, we will ignore them for brevity in the
numerical calculations.  As for $\Delta^{++}$, we have
$F_1=e_\Delta$ and $F_2=\mu_{\Delta^{++}}=(3.7\sim8.5)\mu_N$ from
the average value of the experimental
data~\cite{Nakamura:2010zzi}. From the theoretical calculations,
such as the model-independent way of the chiral quark soliton
model ($\chi$QSM)~\cite{Kim:2003ay,Yang:2004jr} and the SU(6)
quark model~\cite{Beg:1964nm}, it was estimated as
$(5.34\sim5.40)\mu_N$ and $5.58\mu_N$, respectively. Hence, the
middle value from Ref.~\cite{Nakamura:2010zzi}, $5.6\mu_N$ must be
a reasonable choice for numerical calculations,  which leads to
$\kappa_{\Delta^{++}}=3.6$. As for the neutral $\Delta$, we employ
$\kappa_{\Delta^0}=-0.063$~\cite{Nakamura:2010zzi}. The coupling
strength of $g_{\pi N\Delta}$ can be computed  from the
experimental data of its full decay width $\Gamma_{\Delta\to\pi
N}=(116\sim120)$ MeV with $\Gamma_{\Delta}/\Gamma_{\Delta\to\pi
N}\approx100\%$~\cite{Amsler:2008zzb}. By using the Yukawa vertex,
defined by $\mathcal{L}_{PN\Delta}$ in Eq.~(\ref{eq:LAG}), we
obtain the following relation~\cite{Niskanen:1981pq}:
\begin{equation}
\label{eq:decay}
\Gamma_{\Delta\to\pi N}=
\frac{1}{6}\left[\frac{(M_{\Delta}+M_{N})^{2}-m^{2}_{\pi}}{M^{2}_{\Delta}} \right]
\frac{g^{2}_{\pi N\Delta}}{4\pi}\frac{|\bm{p}_{\pi N}|^{3}}{m^{2}_{\pi}},
\end{equation}
where $\bm{p}_{\pi N}$ indicates  the three momentum of the
decaying particle which can be easily calculated by the K\"allen
function~\cite{Amsler:2008zzb}:
\begin{equation}
\label{eq:kollen}
{\bm p}_{\pi N}=\frac{\sqrt{[M^{2}_{\Delta}-(M_{P}+M_{N})^{2}]
[M^{2}_{\Delta}-(M_{P}-M_{N})^{2}]}}{2M_{\Delta}}\approx227\,\mathrm{MeV}.
\end{equation}
Substituting the experimental  information, $m_{\pi}\approx138$
MeV, $M_{N}\approx939$ MeV, and $M_{\Delta}\approx1232$ MeV, into
Eq.~(\ref{eq:decay}) and using Eq.~(\ref{eq:kollen}), one is led
to $g_{\pi N\Delta}\approx(2.14\sim2.18)$. In numerical
calculations, we will make  use of $g_{PN\Delta}=2.16$ as a trial.
As for the $\rho$-meson exchange, the value of $g_{\gamma\pi\rho}$
can be estimated by using the interaction Lagrangian in
Eq.~(\ref{eq:LAG}) for the experimental data
$\Gamma_{\rho^{\pm}\to\gamma\pi^{\pm}}\approx68.59$ keV, and we
have it as $0.245/\mathrm{GeV}$ for the charged decay. The
$g^{(1,2,3)}_{VN\Delta}$ stand for the relevant strong coupling
strengths at the vector-nucleon-$\Delta$ vertex. Again,
taking into account the limited information on these couplings, we
will set them zero  as a trial, except for
$g^{(1)}_{VN\Delta}\equiv g_{\rho N\Delta}$. Using a mesonic
model, the value of $g_{\rho N\Delta}$ was determined as
$(3.5\sim7.8)$ from Ref.~\cite{Flender:1992xw} and references
therein. We will use the average value for it, $g_{\rho
N\Delta}=5.65$ for the numerical calculations.

Using the interaction Lagrangians defined in Eq.~(\ref{eq:LAG})
for the $\gamma p\to\pi^-\Delta^{++}$ and $\gamma
p\to\pi^+\Delta^0$ reaction process, we construct the gauge
invariant amplitudes for the $s$-, $u$-, and $t$-channel
contributions:
\begin{eqnarray}
\label{eq:AMP} i\mathcal{M}_{s}&=&\frac{g_{\pi N\Delta}}{M_{\pi}}
\bar{u}_{\mu}k^{\mu}_{2}\left[
\frac{e_{N}\left[\rlap{/}{k}_{1}F_{s}
+(\rlap{/}{p}_{1}+M_{N})\hat{F}\right]}{s-M^{2}_{N}}
+\frac{e\kappa_{N}}{2M_{N}}
\frac{(\rlap{/}{k}_{1}+\rlap{/}{p}_{1}+M_{N})
F_{s}\rlap{/}{k}_{1}}{s-M^{2}_{N}}\right]\rlap{/}{\epsilon}u,
\cr i\mathcal{M}_{u}&=& \frac{g_{\pi N\Delta}}{M_{\pi}}
\bar{u}_{\mu}\rlap{/}{\epsilon} \left[\frac{
e_{\Delta}[(\rlap{/}{p}_{2}+M_{\Delta})\hat{F}-\rlap{/}{k}_{1}F_{u}]}
{u-M^{2}_{\Delta}} +\frac{e\kappa_{\Delta}}{2M_{\Delta}}
\frac{\rlap{/}{k}_{1}(\rlap{/}{p}_{2}-\rlap{/}{k}_{1}+M_{\Delta})F_{u}}
{u-M^{2}_{\Delta}} \right]G^{\mu\nu}k_{2\nu}u,
\cr i\mathcal{M}_{c} &=&-\frac{e_{\pi}g_{\pi N\Delta}}{M_{\pi}}
\hat{F}\,\bar{u}_{\mu}\epsilon^{\mu}u,
\cr i\mathcal{M}_{t(\pi)}&=& -\frac{2e_{\pi}g_{\pi
N\Delta}}{M_{\pi}}\hat{F}\,
\bar{u}_{\mu}\left(k^{\mu}_{1}-k^{\mu}_{2}\right)
(k_{2}\cdot\epsilon)u\,\mathcal{D}_\pi, \cr
i\mathcal{M}_{t(\rho)}&=& \frac{ig_{\gamma{\pi}\rho}g_{\rho
N\Delta}}{M_{\rho}} F_{t}\,\bar{u}^{\mu}_{2}\gamma_{\nu}
\left[(k^{\mu}_{1}-k^{\mu}_{2})g^{\nu\sigma}-
(k^{\nu}_{1}-k^{\nu}_{2})g^{\mu\sigma}\right]
(\epsilon_{\rho\eta\xi\sigma}k^{\rho}_{1}
\epsilon^{\eta}k^{\xi}_{2})\gamma_5u_1\,\mathcal{D}_\rho,
\end{eqnarray}
where $k_1$, $k_2$, $p_1$, and $p_2$ are the momenta of the
incident photon, outgoing pion, target proton, and recoiled
$\Delta$. $\mathcal{D}_{\pi,\rho}=1/(t-M^2_{\pi,\rho})$ are the
$t$-channel $\pi$  and $\rho$-meson propagators, respectively.
$F_{s,u,t}$ stand for the phenomenological hadronic form factors to consider the spatial distributions of the
relevant hadrons:
\begin{equation}
\label{eq:form}
F_{x}=\frac{\Lambda^{4}}
{\Lambda^{4}+\left(x-M^{2}_{x} \right)^{2}},
\end{equation}
where the subscript $x$ stands for the  Mandelstam
variables and $M_x$ the mass of the off-shell hadron in the $x$ channel~
\cite{Haberzettl:1998eq,Davidson:2001rk,Haberzettl:2006bn}.
The cutoff mass $\Lambda$ will be determined to reproduce the
experimental data in the next Section. $G_{\mu\nu}$ denotes the
projection operator for the spin-$3/2$ fermion, assigned as
\begin{equation}
\label{eq:GG}
G_{\mu\nu}=g_{\mu\nu}-\frac{1}{3}\gamma_{\mu}\gamma_{\nu}
-\frac{2}{3M^{2}_{\Delta}}q_{\mu}q_{\nu}
+\frac{q_{\mu}\gamma_{\nu}-q_{\nu}\gamma_{\mu}}{3M_{\Delta}},
\,\,\,\,q=p_{2}-k_{1}.
\end{equation}
Although the spin-$3/2$ projector is defined as above, we will
simplify it by setting $G_{\mu\nu}\approx g_{\mu\nu}$, which
is verified that no significant distinction is observed by
this simplification as long as we are interested in the two-body
final-state reaction process. Following the prescription
suggested and employed in
Refs.~\cite{Nam:2005uq,Nam:2009cv,Nam:2010au,Haberzettl:1998eq,
Davidson:2001rk,Haberzettl:2006bn}, we employ an overall
form-factor $\hat{F}$ to maintain the WT identity of the
scattering amplitude,
\begin{equation}
\label{eq:fc}
\hat{F}=1-(1-F_{s})(1-F_{t})(1-F_{u}).
\end{equation}
We note that this form-factor scheme  preserves the gauge
invariance, crossing symmetry, and on-shell condition of the form
factors, simultaneously. The electric charges  in
Eq.~(\ref{eq:AMP}) are assigned as
$(e_{\pi},e_{N},e_{\Delta})=(-e,+e,+2e)$ and $(+e,+e,0)$ for the
$\pi^-$ and $\pi^+$ photoproductions, respectively,  to
satisfy the electric-charge conservation.

As mentioned previously, since we are  interested in the region
beyond that of the resonances, it is necessary to take into
account a method to go over the Born approximation which is
believed to be reliable only for the low energy region. As in
Ref.~\cite{Nam:2010au,Regge:1959mz,Vanderhaeghen:1997ts,Yu:2011zu,Yu:2011oo},
the  prescription of the Regge-trajectory for the $t$-channel
meson pole is one of the most successful and practical method for
this purpose. As for  the case of  the $\gamma p\to
K^+\Lambda(1520)$ photoproduction~\cite{Nam:2010au}, where
threshold energy is about $E_\gamma\approx1.67$ GeV,  the Regge
contribution becomes significant beyond $E_\gamma\approx4$ GeV.
Furthermore, the Regge contribution turns out to be responsible
for reproducing the angular dependence of the scattering process
correctly, especially for the momentum-transfer $t$ dependence,
i.e. $d\sigma/dt$ in the high-energy region.

In the present work, we will consider  the Regge trajectories for
the pion and rho meson in the $t$-channel. However, for
simplicity, we will not consider the axial-vector and tensor
mesons, such as the $a_1(1260,1^+)$, $b_1(1235,1^+)$, and
$a_2(1320,2^+)$, and so on. As discussed in
Refs.~\cite{Nam:2010au,Yu:2011zu}, the prescription for the
Reggeization can be applied to the present framework  by replacing
the $t$-channel Feynman propagators in the invariant amplitudes in
Eq.~(\ref{eq:AMP}) with the following one:
\begin{equation}
\label{eq:REGGE} \mathcal{P}_{X}
=\frac{\pi\alpha'_{X}}{\Gamma[\alpha_{X}(t)-J_{X}+1]
\sin[\pi\alpha_{X}(t)]}\left(\frac{s}{s_{0}}
\right)^{\alpha_{X}(t)-J_{X}} ,\,\,\,\, \alpha_{X}(t)=
J_{X}+\alpha'_{X}(t-m^{2}_{X})\mathrm{GeV}^{-2},
\end{equation}
where $\alpha_X$ denotes the Regge  trajectory for the meson $X$
as a function of $t$ with the slope $\alpha'_X$. $J_X$ and $m_X $
stand for the spin and mass of the meson, respectively. Here is a
caveat; in deriving Eq.~(\ref{eq:REGGE}), all the even and odd
spin trajectories are assumed to be degenerate, although in
reality these trajectories are not
degenerated~\cite{Yu:2011zu,Yu:2011oo,Corthals:2005ce,Corthals:2006nz}.
Moreover, for convenience, we have set the phase factor for the
propagators to be positive unity as done in
Ref.~\cite{Ozaki:2009wp}. The cutoff parameter $s_{0}$ is chosen
to be $1$ GeV
conventionally~\cite{Vanderhaeghen:1997ts,Yu:2011zu,Yu:2011oo}.
Since the $\Gamma$ function in Eq.(\ref{eq:REGGE}) plays the role
of the form factor in the Feynman propagator to suppress the
divergence of the Regge propagator at the singularities
$\sin[\pi\alpha_X(t)]=0$, we will not consider the form factors
given in Eq. (\ref{eq:AMP}), setting all of them to unity for the
Regge-trajectory calculations. Hereafter, we use a notation
$i\mathcal{M}^{\mathrm{Regge}}$ for the amplitude constructed with
the Regge propagators in Eq.~(\ref{eq:REGGE}). We list the
relevant inputs for the Regge trajectories in Table~\ref{TAB0}.

\begin{table}[h]
\begin{tabular}{c||c|c|c|c}
\hline &$\hspace{0.5cm}J^P\hspace{0.5cm}$
&$\hspace{0.5cm}m_X\hspace{0.5cm}$
&$\hspace{0.5cm}J_X\hspace{0.5cm}$\
&$\hspace{0.5cm}\alpha'_X\hspace{0.5cm}$\\
\hline \hline
$\hspace{0.2cm}\pi\hspace{0.2cm}$&$0^-$&$140$ MeV&$0$&$0.7$\\
$\hspace{0.2cm}\rho\hspace{0.2cm}$&$1^-$&$770$ MeV&$1$&$0.8$\\
\hline
\end{tabular}
\caption{Relevant inputs for the  Regge trajectory of the meson
$X$.} \label{TAB0}
\end{table}

Let us now discuss the  gauge invariance of the present invariant
amplitude where the $\pi$ and $\rho$ exchanges in
Eq.~(\ref{eq:AMP}) are Reggeized with the Regge propagators in
Eq.~(\ref{eq:REGGE}). Following the procedure in
Refs.~\cite{Vanderhaeghen:1997ts,Yu:2011zu,Yu:2011oo,Corthals:2005ce,Corthals:2006nz}
for the $t$-channel Reggeization, we write the Reggeized amplitude
for the $\pi$ exchange as
\begin{equation}
\label{eq:WT1}
i\mathcal{M}=i\mathcal{M}^{\mathrm{Regge}}_{t(\pi,\rho)}
+(i\mathcal{M}_{s}+i\mathcal{M}_{u}
+i\mathcal{M}_{c}).
\end{equation}
It is, then, easy to show that the amplitude
in Eq.~(\ref{eq:WT1}) does not satisfy the current conservation (WT identity) as follows:
\begin{equation}
\label{eq:}
 k_1\cdot(i\mathcal{M})=k_{1}\cdot(i\mathcal{M}^\mathrm{Regge}_{t(\pi)}
 +i\mathcal{M}^{E}_{s}
+i\mathcal{M}^{E}_{u}+i\mathcal{M}_{c})\ne0.
\end{equation}
Hence, the sum of the $\pi$-exchange in the $t$-channel and
electric  $s$- and $u$-channel contributions must be zero to
satisfy the gauge invariance, whereas the $\rho$-exchange and
magnetic contributions are automatically zero, due to their
antisymmetric nature. To satisfy the gauge invariance, we follow
the prescription, suggested in
Refs.~\cite{Vanderhaeghen:1997ts,Yu:2011zu,Yu:2011oo,Corthals:2005ce,Corthals:2006nz}
as
\begin{equation}
\label{eq:CHANGE}
i\mathcal{M}^\mathrm{Regge}_{t(\pi)}
 +i\mathcal{M}^{E}_{s}
+i\mathcal{M}^{E}_{u}+i\mathcal{M}_{c}\to
i\mathcal{M}^\mathrm{Regge}_{t(\pi)}
 +(i\mathcal{M}^{E}_{s}
+i\mathcal{M}^{E}_{u}+i\mathcal{M}_{c})(t-m^2_\pi)\mathcal{P}_{\pi}
\equiv i\bar{\mathcal{M}}^\mathrm{Regge},
\end{equation}
where we set $F_{s,u}$ and  $\hat{F}$ to be unity for the
$i\bar{\mathcal{M}}^\mathrm{Regge}$. Thus, the Reggeized amplitude
can be written as follows:
\begin{equation}
\label{eq:REGGEAMP}
i\mathcal{M}=i \bar{\mathcal{M}}^\mathrm{Regge}
+i\mathcal{M}^M_s+i\mathcal{M}^M_u+i\mathcal{M}^\mathrm{Regge}_{t(\rho)},
\end{equation}
where the superscript $M$ denotes the magnetic contributions of the $s$ and $u$ channels.

We recall that the Regge propagators should work
properly for the energy and momentum transfer
$(s,|t|)\to(\infty, 0)$, but in practice the Regge contributions
could nevertheless affect even in the low-energy region
$(s,|t|)\to(s_{\mathrm{threshold}}, \mathrm{finite})$. Thus, it is physically reasonable to suppose that the meson propagators are smoothly shifted from the Regge
one at high energy for $s\gtrsim s_{\mathrm{Regge}}$ to the
usual Feynman one  in the low energy for $s\lesssim
s_{\mathrm{Regge}}$. Here, $s_{\mathrm{Regge}}$ indicates a
certain value of $s$ from which the Regge contributions become
effective. As discussed in Ref.~\cite{Nam:2010au} in detail, there
is no unique scheme to interpolate these two regions. Thus,
as a trial, we introduce to the relevant invariant amplitudes
in Eq.~(\ref{eq:AMP}) an ansatz which interpolates between the
Feynman and Regge realms by redefining the form factors as
follows:
\begin{equation}
\label{eq:R} \hat{F}\to\bar{F}_{c}\equiv
\left[(t-M^{2}_{X})\mathcal{P}_{X}\right]
\mathcal{R}+\hat{F}(1-\mathcal{R}),
\end{equation}
where the ansatz for the interpolation reads
\begin{equation}
\label{eq:RSRT} \mathcal{R}=\frac{1}{2}
\left[\mathrm{tanh}\left(\frac{s-s_{\mathrm{Regge}}}{s'}\right)+1\right].
\end{equation}
Here, $s'$ denotes a free parameter to make the argument of
$\tanh$ in Eq.~(\ref{eq:RSRT}) dimensionless. It is easy to
understand that $\mathcal{R}$ goes to unity as $s\to\infty$ and
approaches to zero as $s\to 0$. This asymptotic behaviors of
$\mathcal{R}$ ensures that $\bar{F}_{c}$ in Eq.~(\ref{eq:R})
interpolates the two energy regions smoothly. As already shown in
Ref.~\cite{Nam:2010au}, this interpolation could describe
qualitatively very well the low- and high-energy region data,
simultaneously. We will determine the parameters,
$s_{\mathrm{Regge}}$ and $s'$, with experimental data in the next
Section. We want to   indicate the difference of the
interpolation function in the present work from that in
Ref.~\cite{Nam:2010au} briefly. In the previous work, there was
an additional term which was multiplied to
Eq.~(\ref{eq:RSRT}) as a function of $t$, as we considered the
$t$ dependence of the Regge trajectories. However, since we
verified that it is not effective for a considerably wide energy
range we will ignore such a $t$-dependent term here, and as a
result, we have the interpolation function as a function of $s$ as
in Eq.~(\ref{eq:RSRT}). Consequently, as discussed above, we will
have three distinctive models in the following Sections, assigned
as:
\begin{enumerate}
\item {\it Born}: The scattering amplitude is defined with the conventional Feynman  propagators with the phenomenological form factors as in Eqs.~(\ref{eq:AMP}), (\ref{eq:form}), and (\ref{eq:fc}).
\item {\it Regge}: The scattering amplitude is modified by the Regge contributions without form factors as in Eq.~(\ref{eq:REGGEAMP}).
\item {\it Interpolation}: The {\it Regge} approach is modified by the interpolation formula and ansatz in Eqs.~(\ref{eq:R}) and (\ref{eq:RSRT}).
\end{enumerate}

\section{Numerical results and Discussions}

In this Section, we present  the numerical results for the various
physical observables such as the energy and angular dependences of
the cross sections for the two charged pion photoproduction processes.
We also show the numerical results for the polarization-transfer coefficient
($C_{x,z}$)~\cite{McNabb:2003nf,Anisovich:2007bq,Nam:2009cv} and the
$\pi^{+}$-decay angle distribution~\cite{Barber:1980zv,Nam:2010au}
for the two production processes. The results are
compared with experimental data, and we provide theoretical estimations useful for future experiments.

\begin{figure}[ht]
\includegraphics[width=12cm]{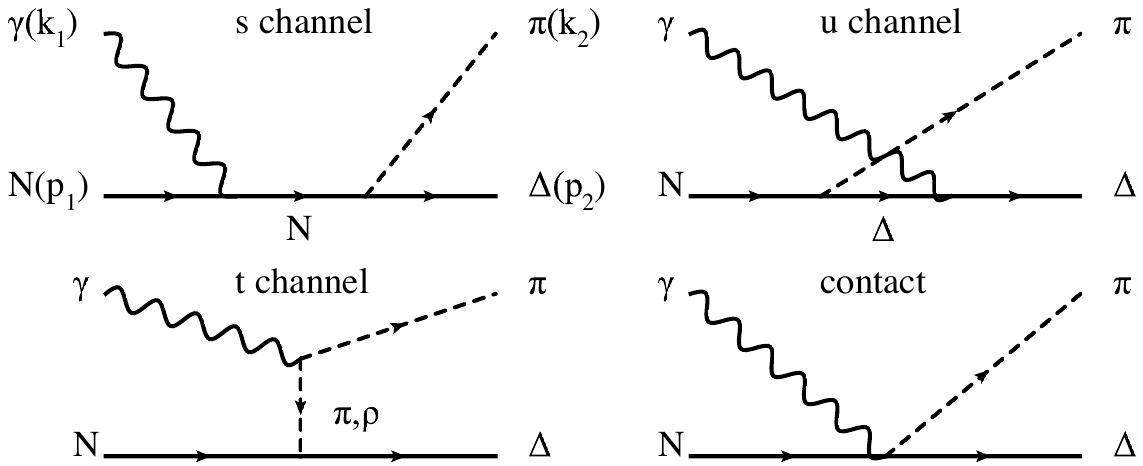}
\caption{Relevant Feynman diagrams for the $\gamma N\to \pi\Delta(1232)$ reaction process.}
\label{FIG0}
\vspace{1cm}
\begin{tabular}{cc}
\includegraphics[width=8.5cm]{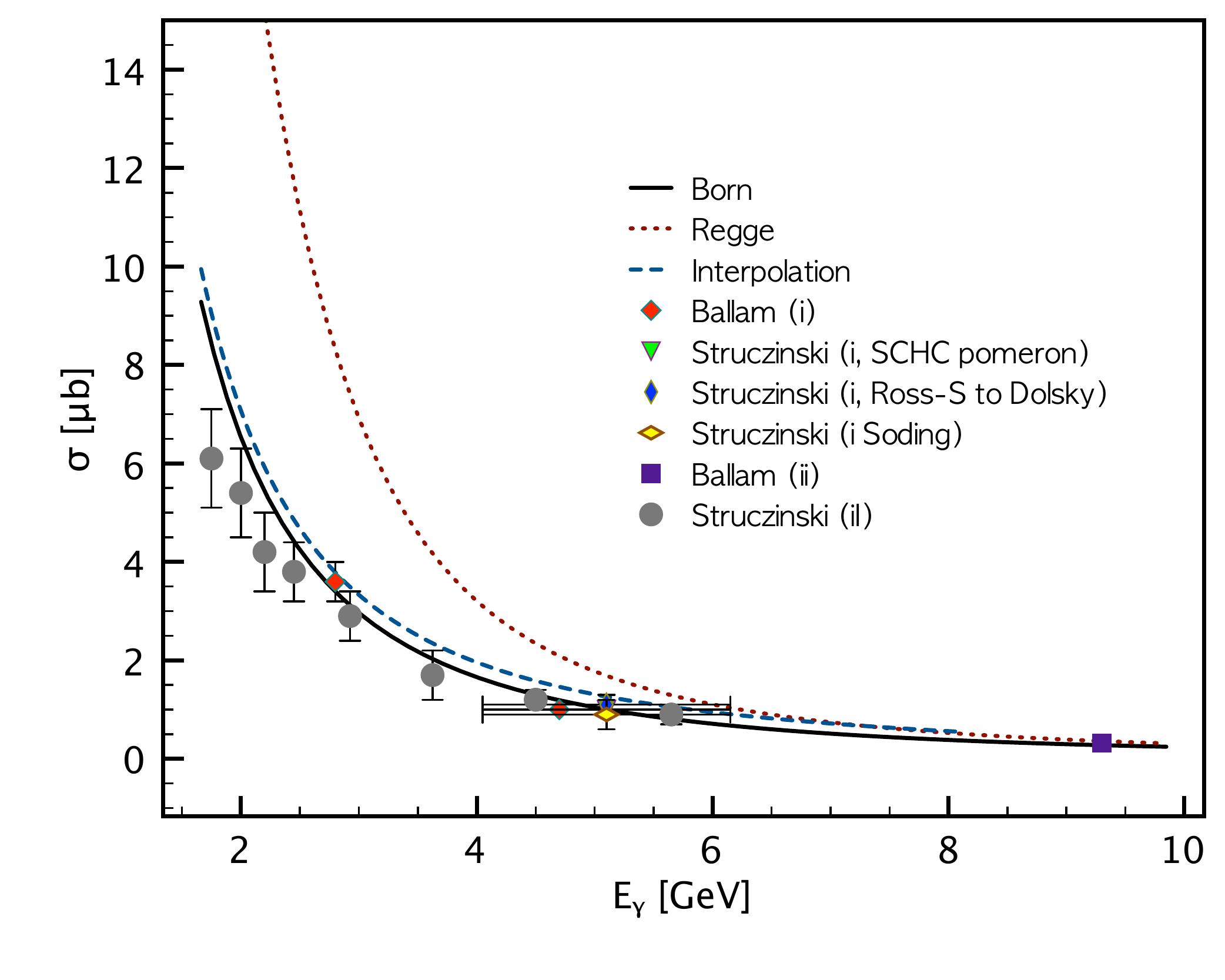}
\includegraphics[width=8.5cm]{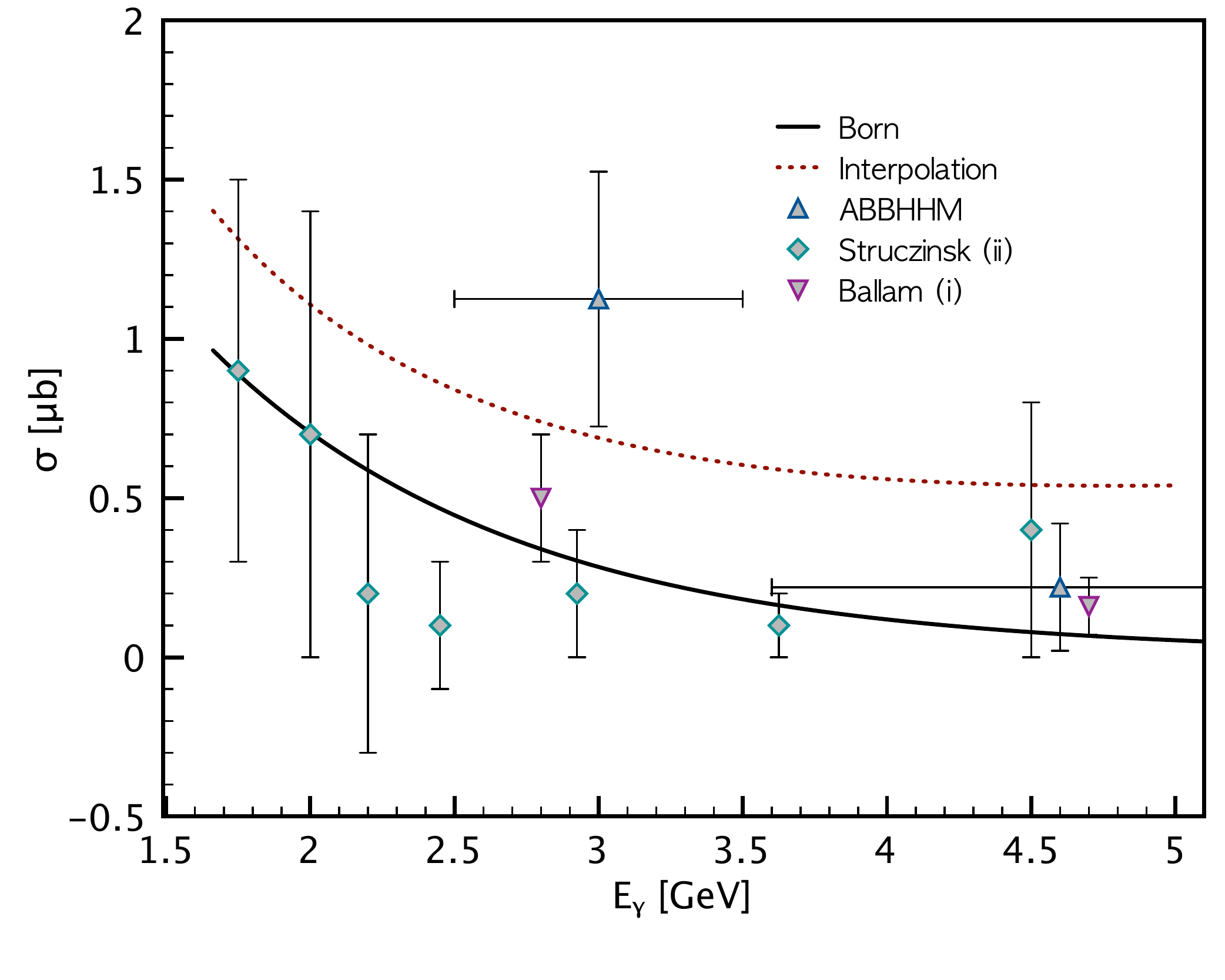}
\end{tabular}
\caption{(Color online) Total cross sections for the $\gamma p\to\pi^{-}\Delta^{++}$ (left) and $\gamma p\to\pi^{+}\Delta^{0}$ reaction processes as a function of $E_{\gamma}$. Experimental data are taken from Refs.~\cite{Ballam:1971yd} (Ballam (i)), \cite{Struczinski:1973we} (Struczinski (i)), \cite{Ballam:1972eq} (Ballam (ii)), \cite{Struczinski:1975ik} (Struczinski (ii)), and \cite{BrownHarvardMITPadovaWeizmannInstituteBubbleChamberGroup:1967zz} (ABBHHM). We show the curves from the numerical results for the Born terms (solid),  that plus the Regge (dot), those with the interpolation (dash), separately.  For more details, see the text.}
\label{FIG1}
\vspace{1cm}
\begin{tabular}{cc}
\includegraphics[width=8.5cm]{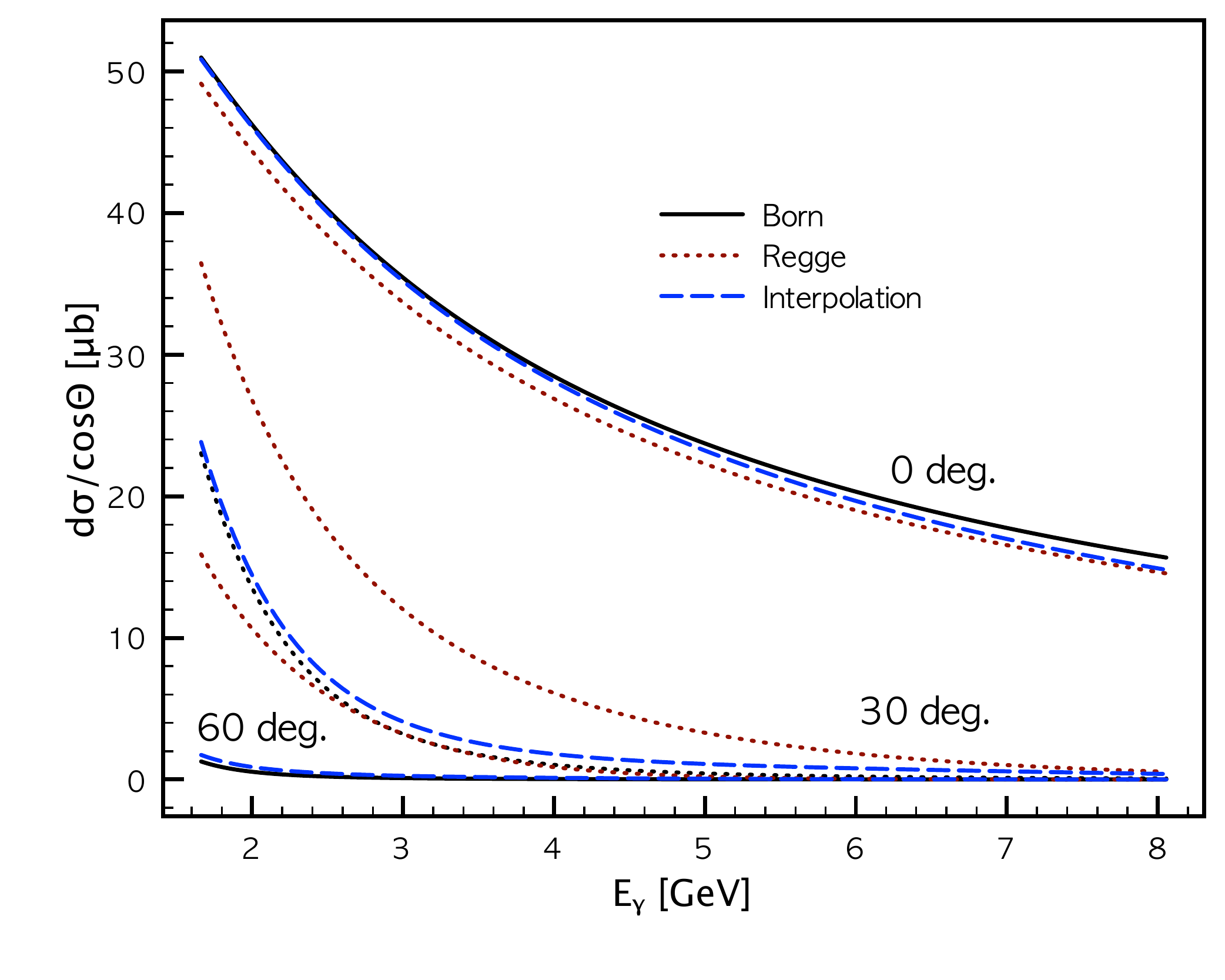}
\includegraphics[width=8.5cm]{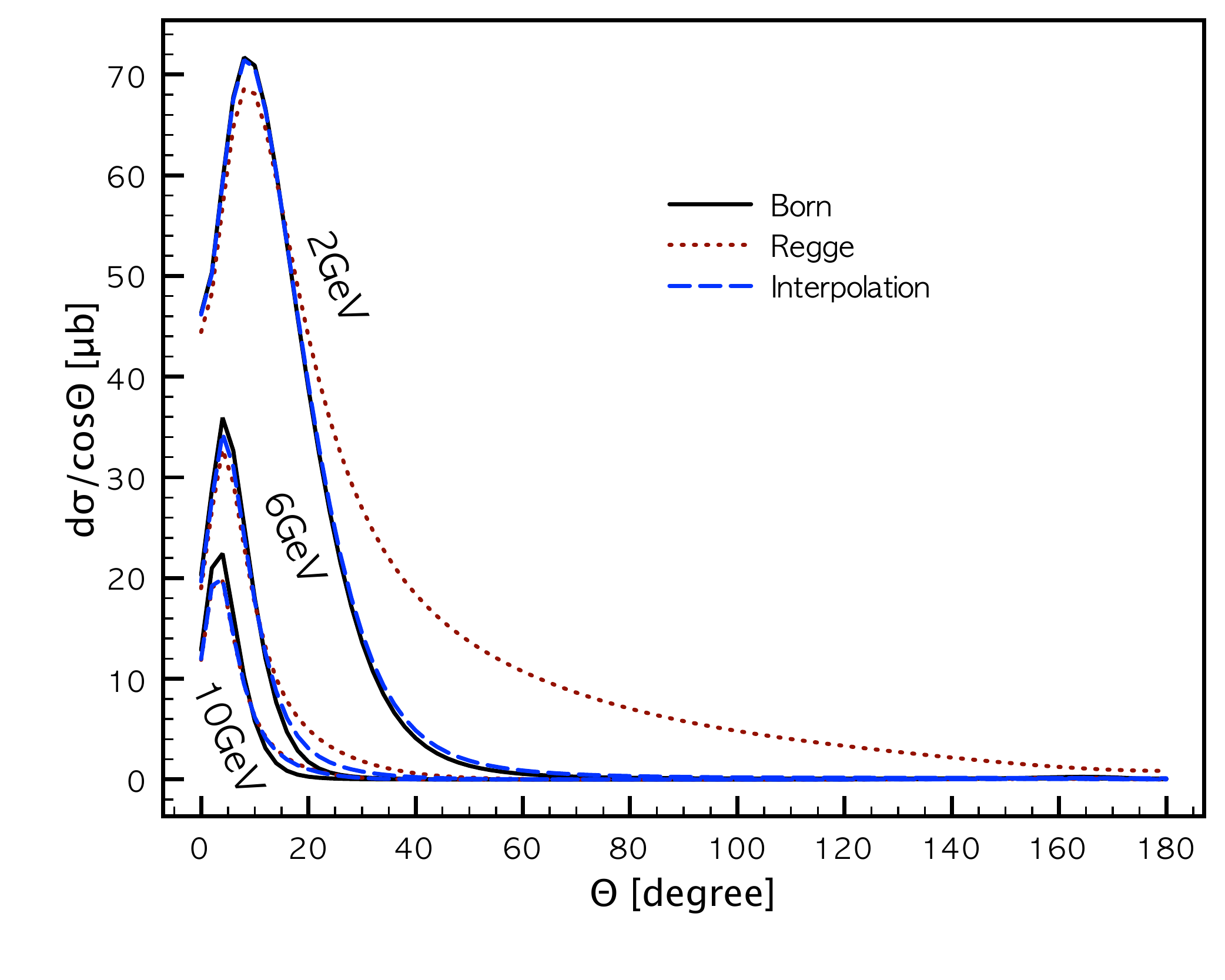}
\end{tabular}
\caption{(Color online) Differential cross section for $\gamma p\to\pi^{-}\Delta^{++}$ as a function of $E_{\gamma}$ for different angles $\theta=(0,30^{\circ},60^{\circ})$ for the Born terms (solid),  that plus the Regge (dot), those with the interpolation (dash), separately in the left panel. In the right panel, we plot the differential cross section as a function of $\theta$ for different energies, $E_{\gamma}=(2,6,10)$ GeV, represented in the same manner with the left panel.}
\label{FIG2}
\end{figure}

First, we investigate the unpolarized total cross section as a
function of the photon energy $E_{\gamma}$ in the three different
models. We present the results for the $\pi^-$ photoproduction in
the left panel of Fig.~\ref{FIG1} in which the solid, dotted, and
dot-dashed lines correspond to the Born, Regge, and interpolation
models, respectively. The experimental data are taken from
Refs.~\cite{Ballam:1971yd,Struczinski:1973we,Ballam:1972eq}. All
the calculations are performed with the cutoff mass
$\Lambda\approx450$ MeV, determined to reproduce the data. Note
that this cutoff mass is about $25\%$ smaller than that for the
$\Lambda(1520)$
photoproduction~\cite{Nam:2005uq,Nam:2009cv,Nam:2010au}. The
simple Born model provides qualitatively good agreement with the
data, whereas the Regge one corresponding to $\mathcal{R}=1$ shows
overshoot in the lower energy region. This observation indicates
that the simple replacement of the $t$-channel Feynman propagators
with the Regge ones in the present case does not work so well. In
the application of the interpolation ansatz with the relevant
parameters determined as
$s_{\mathrm{Regge}}=(3.5\,\mathrm{GeV})^2$ and
$s'=(2.5\,\mathrm{GeV})^2$, we obtain a reasonable result as shown
in Fig.~\ref{FIG1}. These values indicate that the Regge
contribution starts to dominate the present reaction process
beyond $\sqrt{s}\approx3.5$ GeV.  As for the $\pi^+$
photoproduction, it turns out that one needs much smaller cutoff
mass to reproduce the experimental data qualitatively, i.e.
$\Lambda\approx200$ MeV as shown in the right panel of
Fig.~\ref{FIG1} for the Born and interpolation models. Note that
this cutoff value is much smaller than that for the $\pi^-$
photoproduction, signaling that there can be considerable
contributions  from those beyond the ground-state nucleon, for instance. The numerical results for the Born and
interpolation models show similar energy dependence, whereas their
strengths are rather different, although the experimental data
contain sizable uncertainties in this channel. Note that the
Regge-model result strongly overshoots the data which are not
depicted here.

The left panel of Fig.~\ref{FIG2} shows the differential cross
section for the $\pi^-$ photoproduction as a function of
$E_{\gamma}$ for the three different angles,
$\theta=(0,30^{\circ},60^{\circ})$ with the same legends for the
lines with those in Fig.~\ref{FIG1}. At the very forward angle
$\theta=0$, the strength of the cross section decreases
monotonically for the three cases without showing significant
differences. On the contrary, as the angle increases, the
difference between them becomes obvious, according to the
considerable contributions from the Regge trajectories. The
difference between the models with and without the interpolation
turns out to be moderate for $\theta\gtrsim30^{\circ}$, and
becomes negligible at $\theta=60^{\circ}$.

In the right panel of Fig.~\ref{FIG2},  we draw the differential
cross section as a function of $\theta$ for different energies,
$E_{\gamma}=(2,6,10)$ GeV, presented in the same manner with
the left panel. From this we can conclude that there is the strong
$t$-channel contribution, which makes a peak in the forward
scattering region. This tendency is rather different from that of
the $\Lambda(1520)$ photoproduction, in which case the
contact-term contribution dominates the reaction
process~\cite{Nam:2005uq,Nam:2009cv,Nam:2010au}.  As the energy
increases, the overall strength of the differential cross sections
become smaller and the peak position moves to the vicinity around
$\theta=0$. We will discuss the shift of the peak-position in
detail below. We note that the Born and interpolation models provide almost negligible
contributions in the backward scattering region,
$\theta\gtrsim60^{\circ}$, which can be understood easily from the
left panel of Fig.~\ref{FIG2}: the smaller strengths for the larger angles.

\begin{figure}[ht]
\begin{tabular}{cc}
\includegraphics[width=8.5cm]{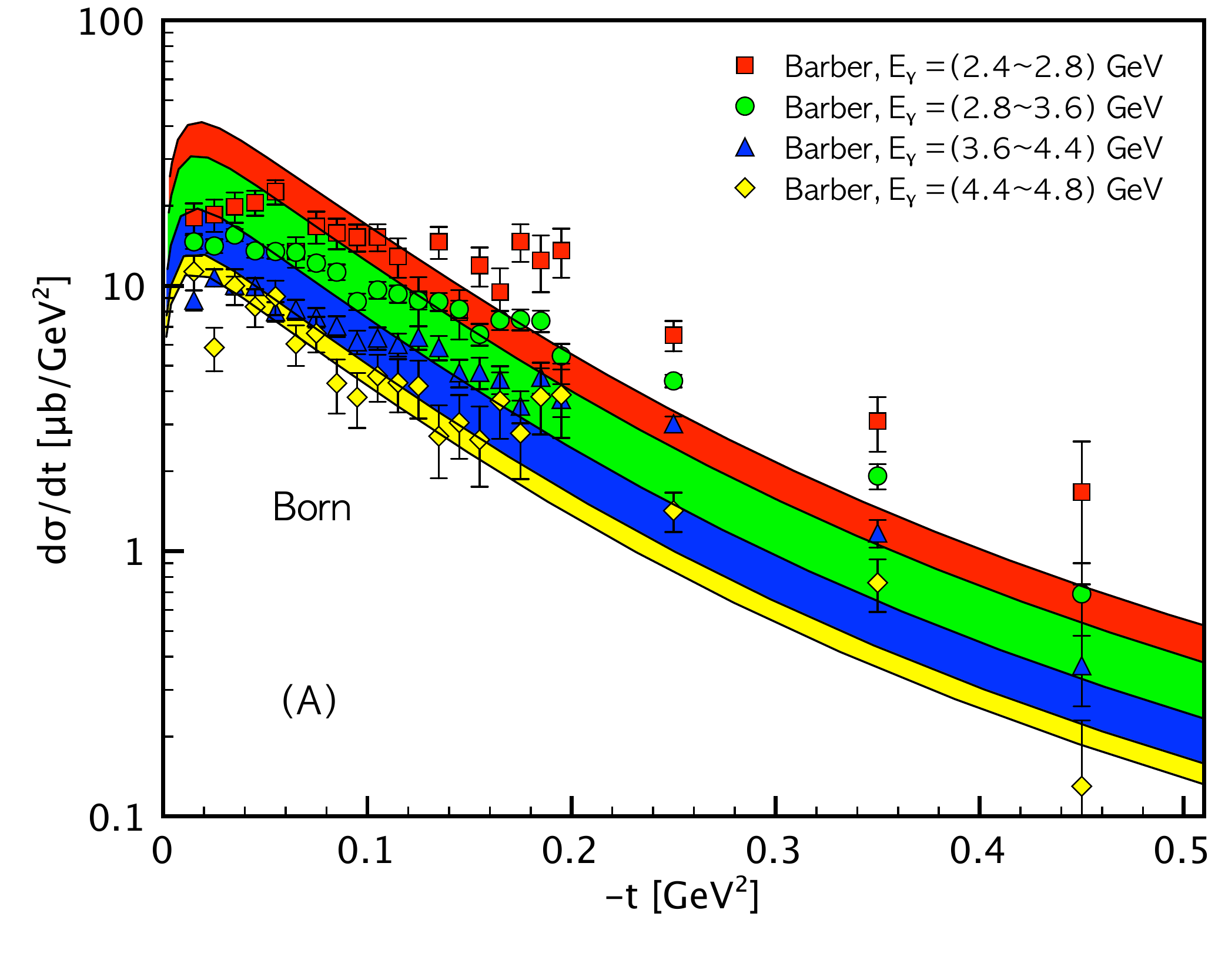}
\includegraphics[width=8.5cm]{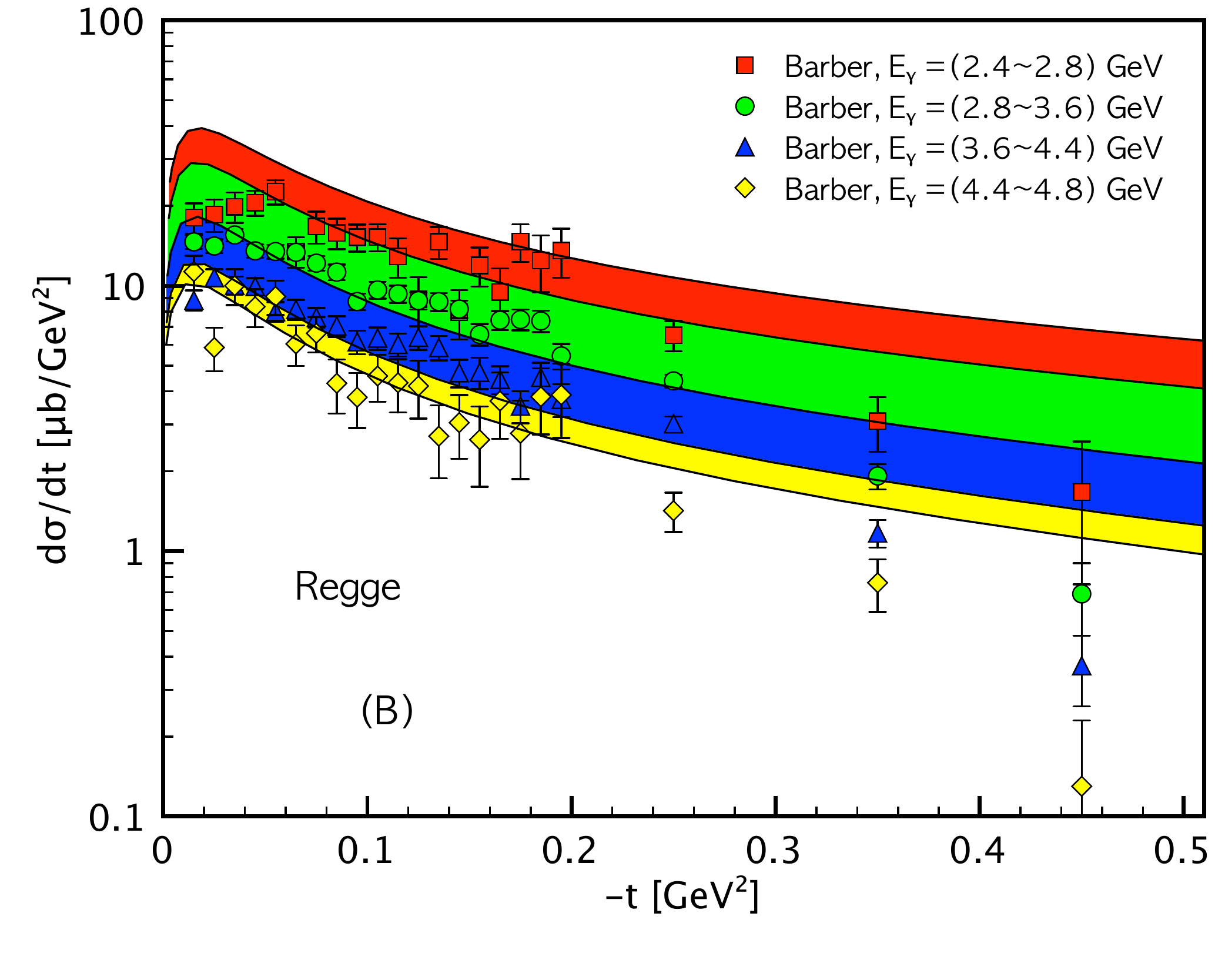}
\end{tabular}
\begin{tabular}{cc}
\includegraphics[width=8.5cm]{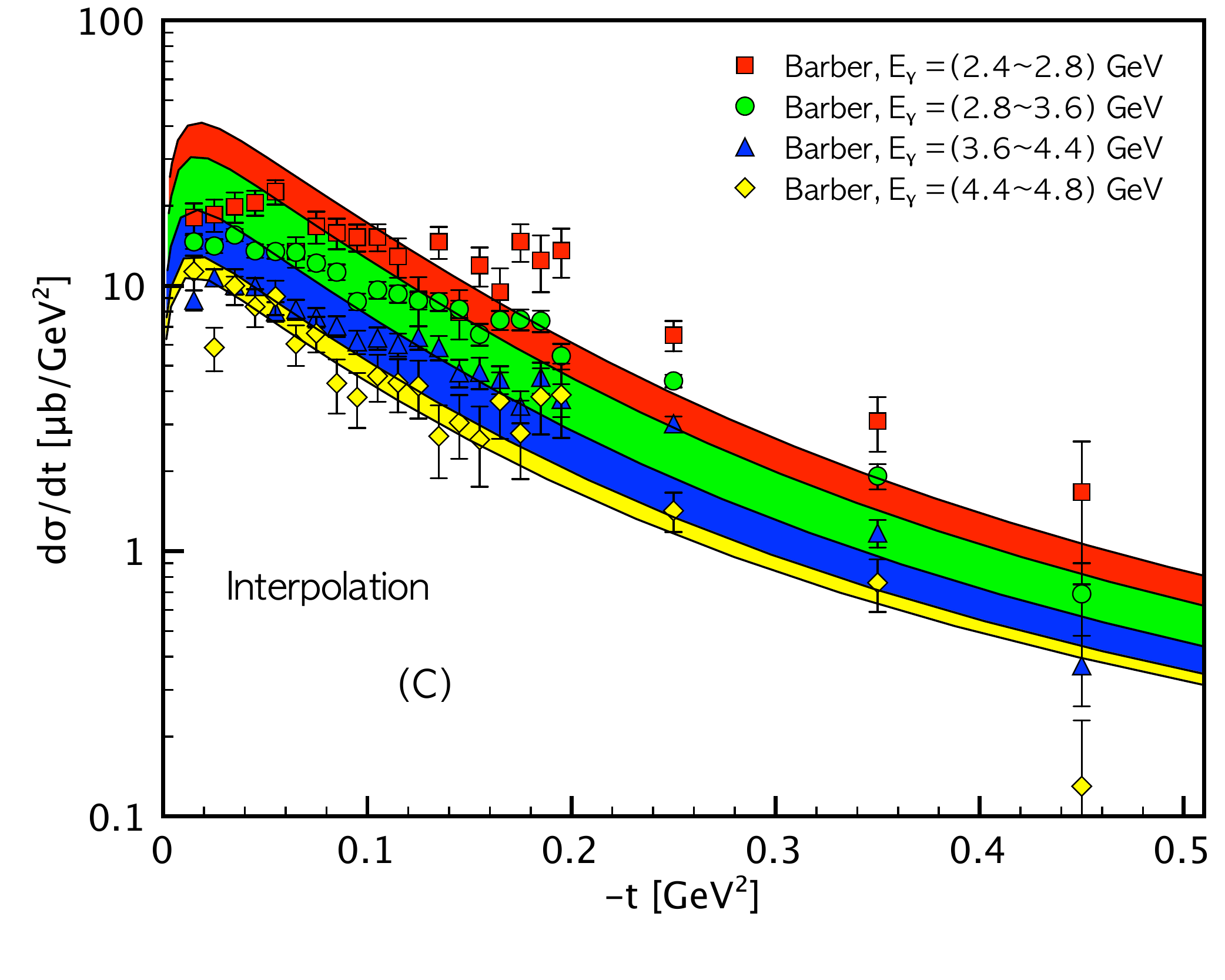}
\includegraphics[width=8.5cm]{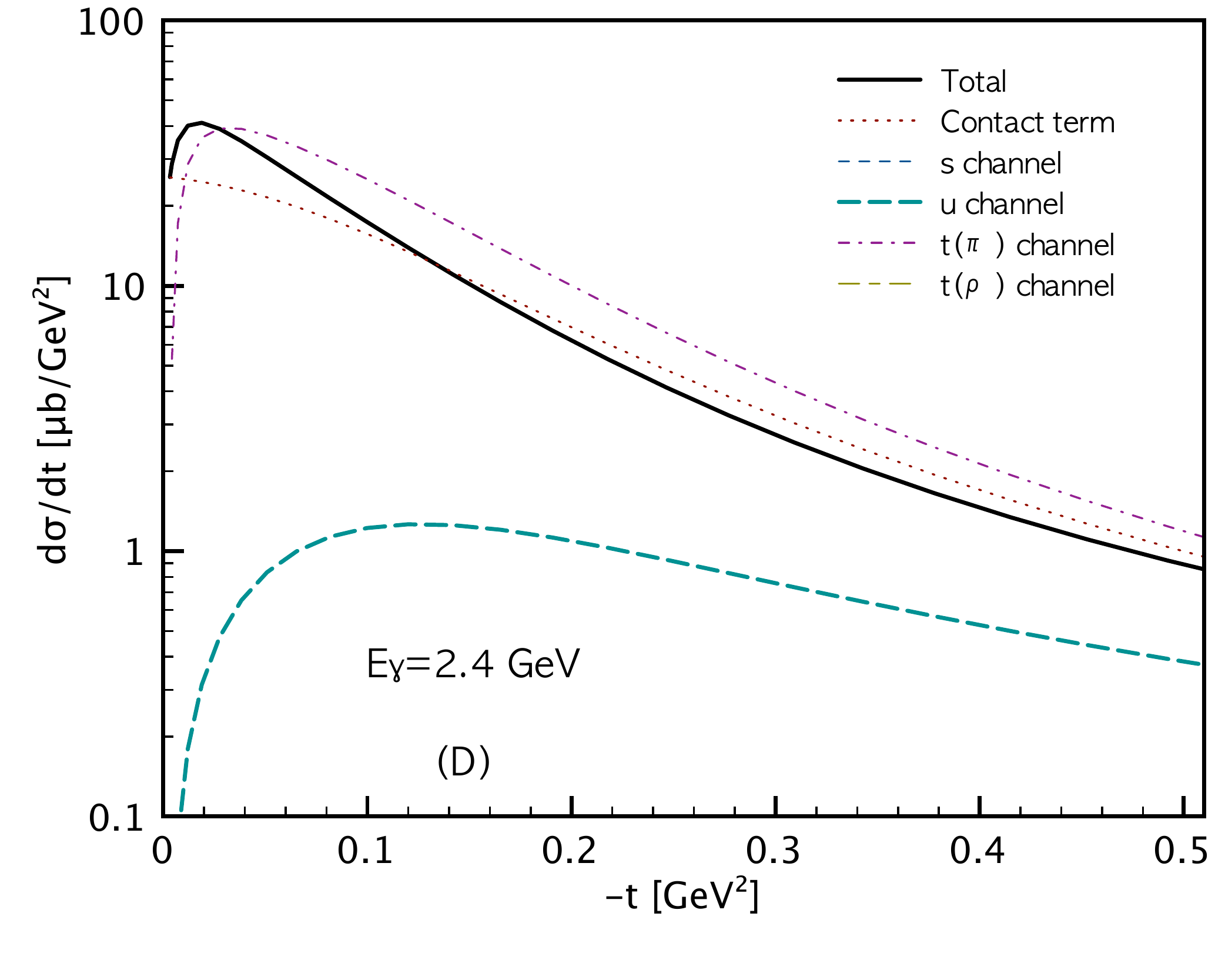}
\end{tabular}
\caption{(Color online) Momentum-transfer dependence $d\sigma/dt$ [$\mu$b/GeV$^{2}$] for $\gamma p\to\pi^{-}\Delta^{++}$ as a function of $-t$ [GeV$^{2}$] for the low-energy region $E_{\gamma}=(2.4\sim4.8)$ GeV for the Born (A), Regge (B), interpolation (C) models, separately. Experimental data are taken from Ref.~\cite{Barber:1980dp} (Barber). Each contribution is also drawn for $E_\gamma=2.4$ GeV for the Born model (D). In the panel (E), we draw the numerical result for the $\gamma p\to \pi^+\Delta^0$, using only the interpolation model.}
\label{FIG3}
\includegraphics[width=8.5cm]{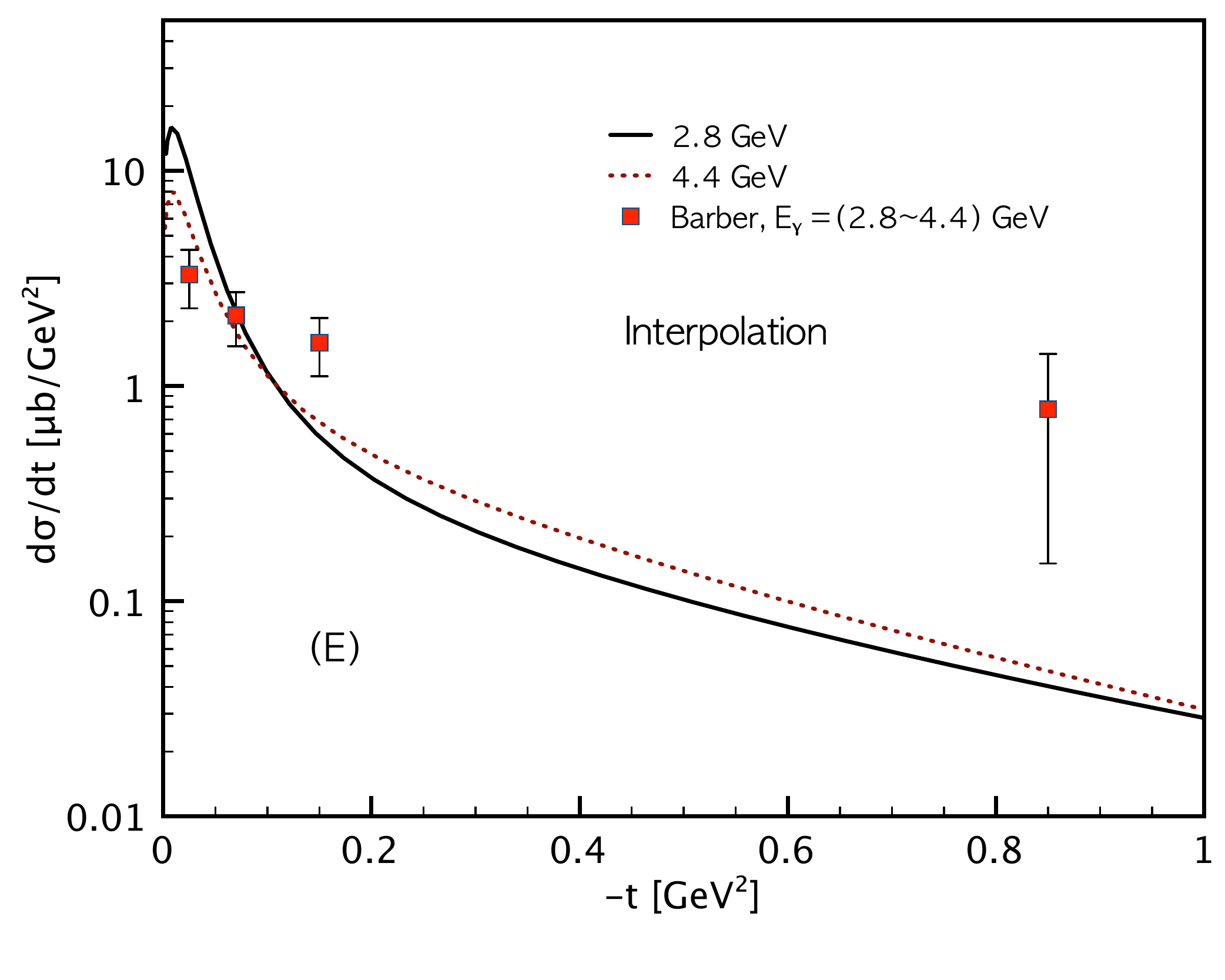}
\caption{(Color online) Momentum-transfer dependence $d\sigma/dt$ [$\mu$b/GeV$^{2}$] for $\gamma p\to\pi^{+}\Delta^{0}$ as a function of $-t$ [GeV$^{2}$] for the low-energy region $E_{\gamma}=(2.4\sim4.8)$ GeV for the interpolation model. Experimental data are taken from Ref.~\cite{Barber:1980dp} (Barber). }
\label{FIG3-1}
\end{figure}

Now, we are in a position to discuss the momentum-transfer dependence of the present reaction process, represented by
$d\sigma/dt$ as a function of $-t$ in Fig.~\ref{FIG3}. The
three cases for the Born (A), Regge (B), and interpolation (C) models are presented separately for the four different energy ranges, $E_{\gamma}=(2.4\sim2.8)$ GeV,
$(2.8\sim3.6)$ GeV, $(3.6\sim4.4)$ GeV, and,
$(4.4\sim4.8)$ GeV, in which the shaded band stands for
each energy interval. The experimental data are taken from
Ref.~\cite{Barber:1980dp}. As for the Born model (A) in Fig.~\ref{FIG3}, the experimental data are qualitatively well reproduced in the region
$-t=(0.05\sim0.2)\,\mathrm{GeV}^{2}$. Outside that region, only the
high energy data around $E_{\gamma}\approx4.6$ GeV are relatively
in good agreement with the theoretical prediction. In other words,
these observations tell us that there can be missing contributions
for the lower energy region in the present framework, especially
for the small $|t|$ region. Such discrepancies indicate that one may need to consider further contributions from the $N$ and $\Delta$ resonances around
$E_\mathrm{cm}=(2\sim3)$ GeV and possibly from other kinds of
$t$-channel meson exchanges, which we are not considering here. As
we take into account the Regge contributions as shown in the panel B, there
appears strong enhancement for the region beyond
$-t\gtrsim0.2\,\mathrm{GeV}^{2}$, which overshoots the data.
Again, this unexpected overestimation is tamed by including the
interpolating ansatz as seen in the panel C of Fig.~\ref{FIG3},
which leads to relatively a good agreement with the experimental
data. In the panel D, we draw each contribution separately for the
$d\sigma/dt$ at $E_\gamma=2.4$ GeV, using the Born model, in order to see which
contribution is essential to produce the curve. As shown there, the
$\pi$-exchange and contact-term contribution dominate the forward
scattering region, whereas the others are considerably small.
Moreover, one can easily see that the contact-term contribution
makes the peak shifted to the smaller $t$ region
($-t\approx0.01\,\mathrm{GeV}^2$). By comparing this observation
with the right panel of Fig.~\ref{FIG2} and seeing the shift of
the peak, one can easily find that the contact-term contribution
gets prevailing over the $\pi$-exchange as $E_\gamma$ increases,
although the contribution of the $\pi$-exchange is still dominant. In the panel E of Fig.~\ref{FIG3-1}, we also show the numerical results, using the interpolation model, for the $\pi^+$ photoproduction. Again, we observe similar tendency with that for the $\pi^-$ one.

\begin{figure}[ht]
\begin{tabular}{cc}
\includegraphics[width=8.5cm]{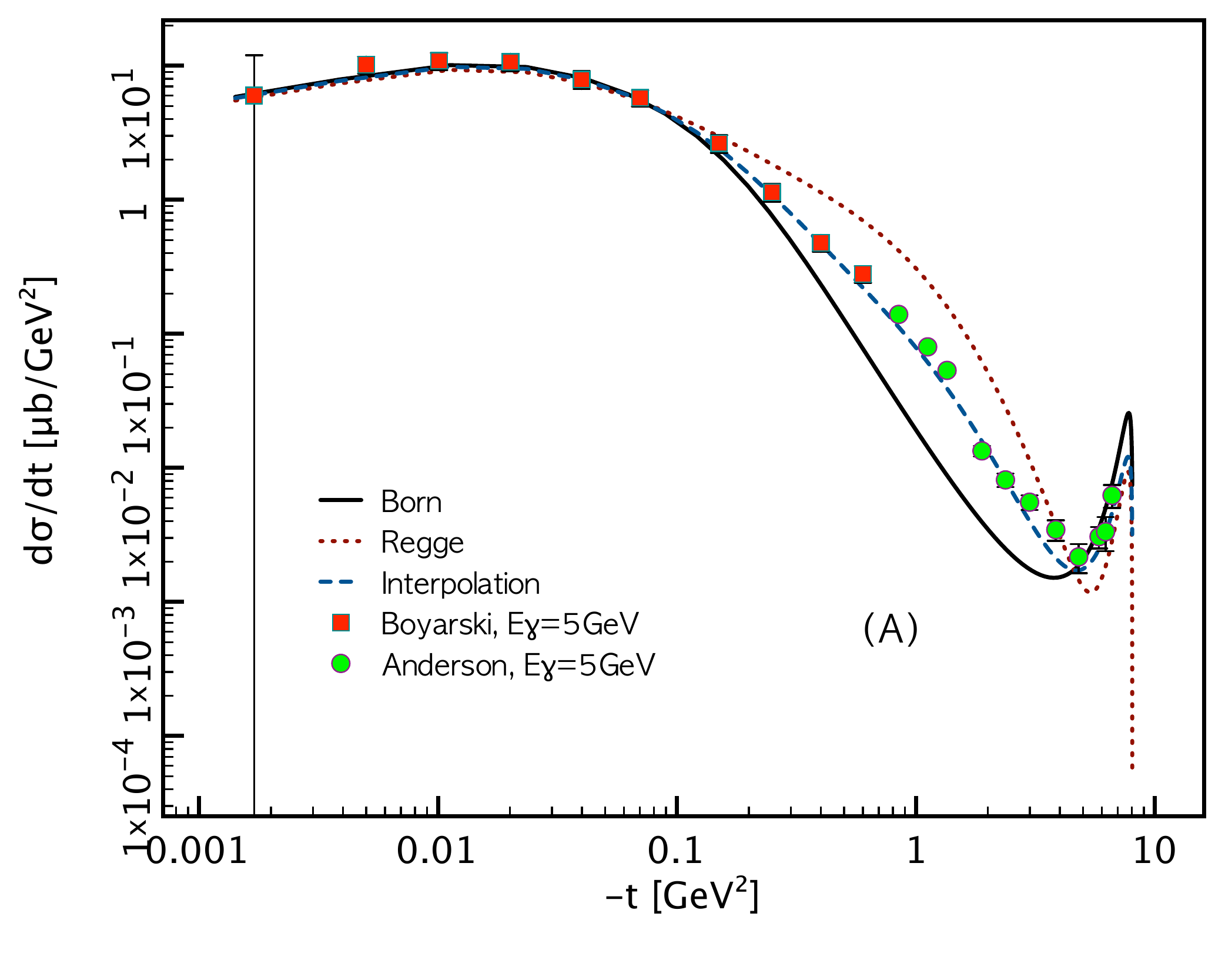}
\includegraphics[width=8.5cm]{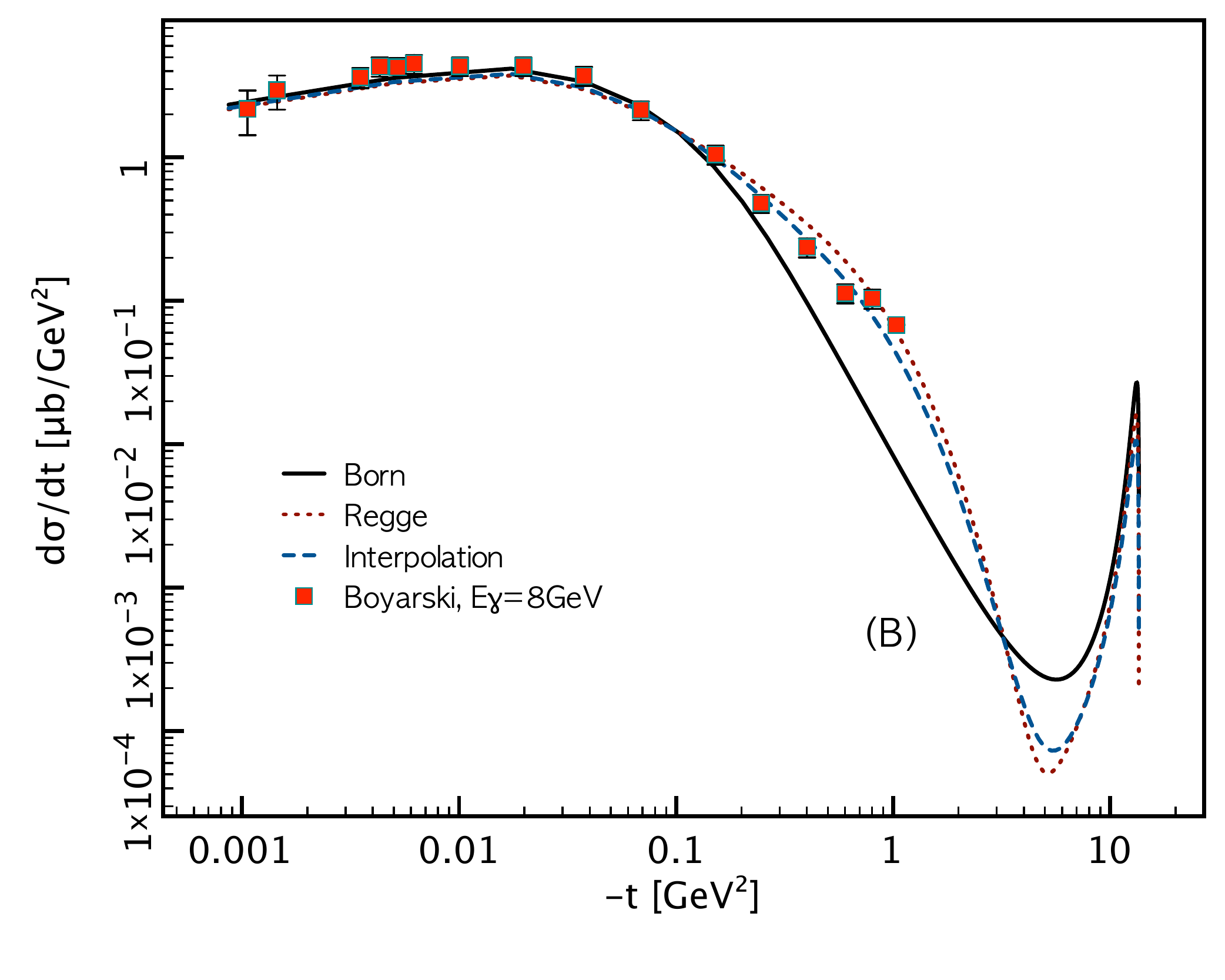}
\end{tabular}
\begin{tabular}{cc}
\includegraphics[width=8.5cm]{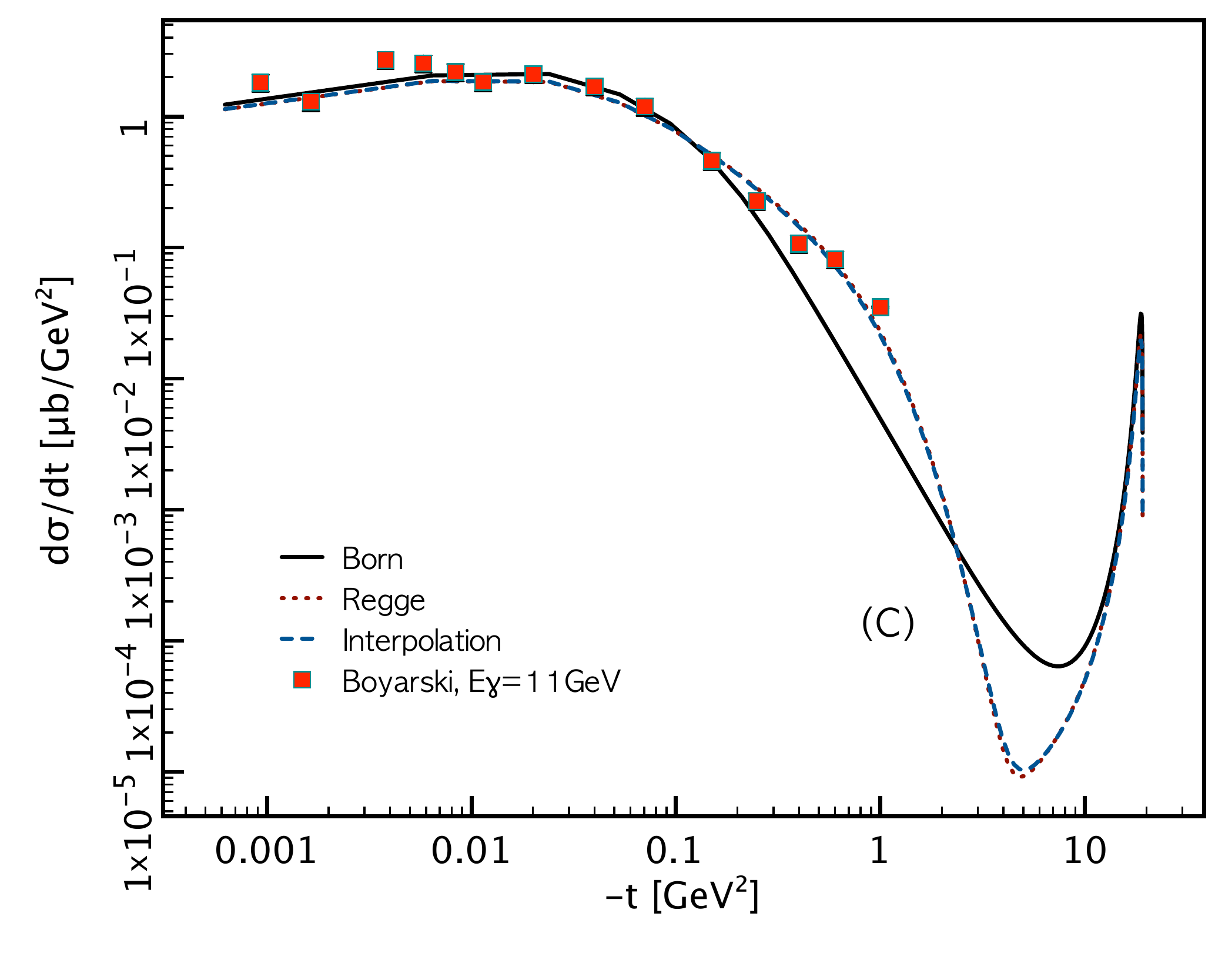}
\includegraphics[width=8.5cm]{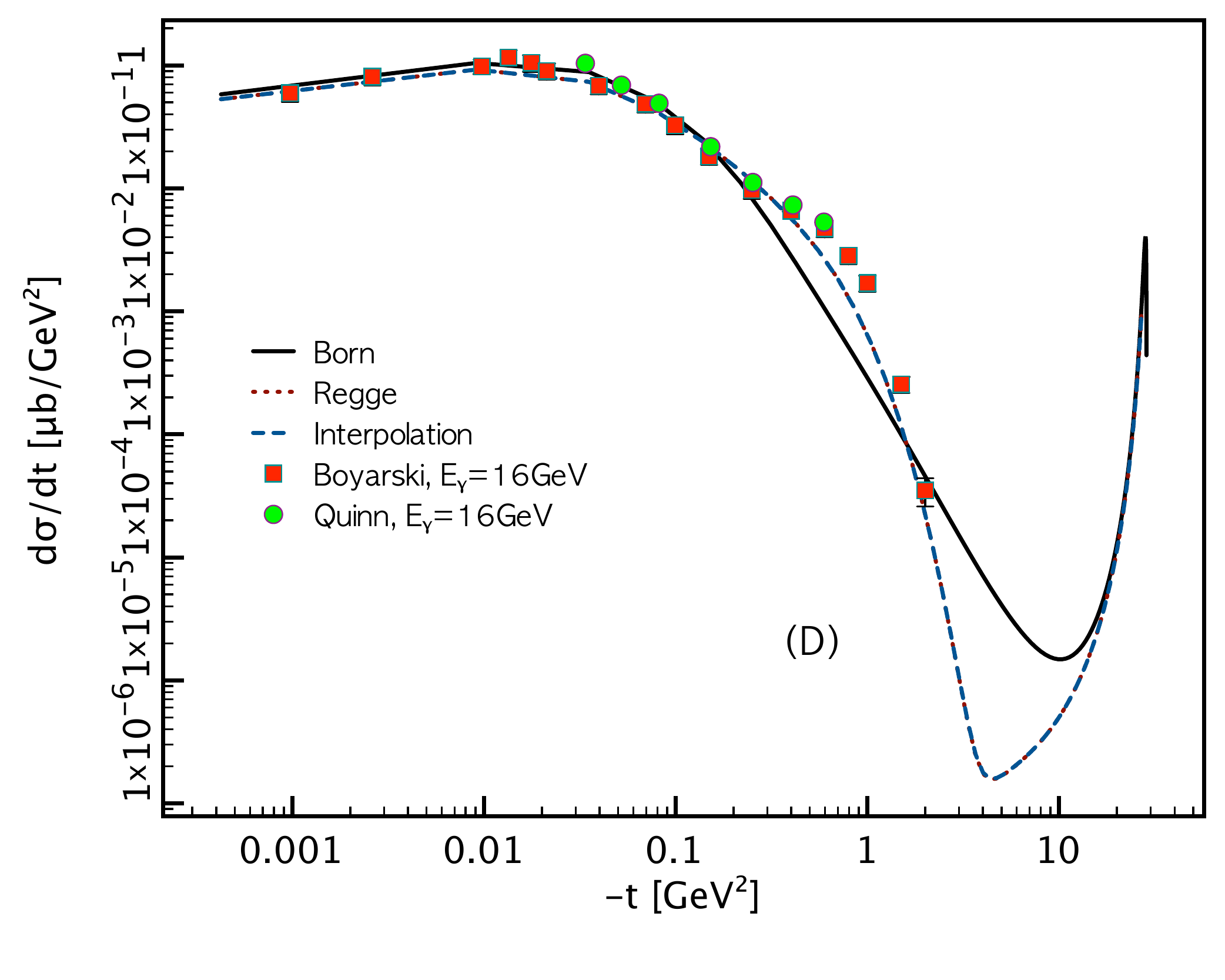}
\end{tabular}
\caption{(Color online) Momentum-transfer dependence $d\sigma/dt$ [$\mu$b/GeV$^{2}$] for $\gamma p\to\pi^{-}\Delta^{++}$ as a function of $-t$ [GeV$^{2}$] for the high-energy region $E_{\gamma}=(5,8,11,16)$ GeV in panel (A,B,C,D). Experimental data are taken from Refs.~\cite{Boyarski:1968dw} (Boyarski),\cite{Anderson:1976ph} (Anderson), and \cite{Quinn:1979zp} (Quinn).}
\label{FIG4}
\includegraphics[width=8.5cm]{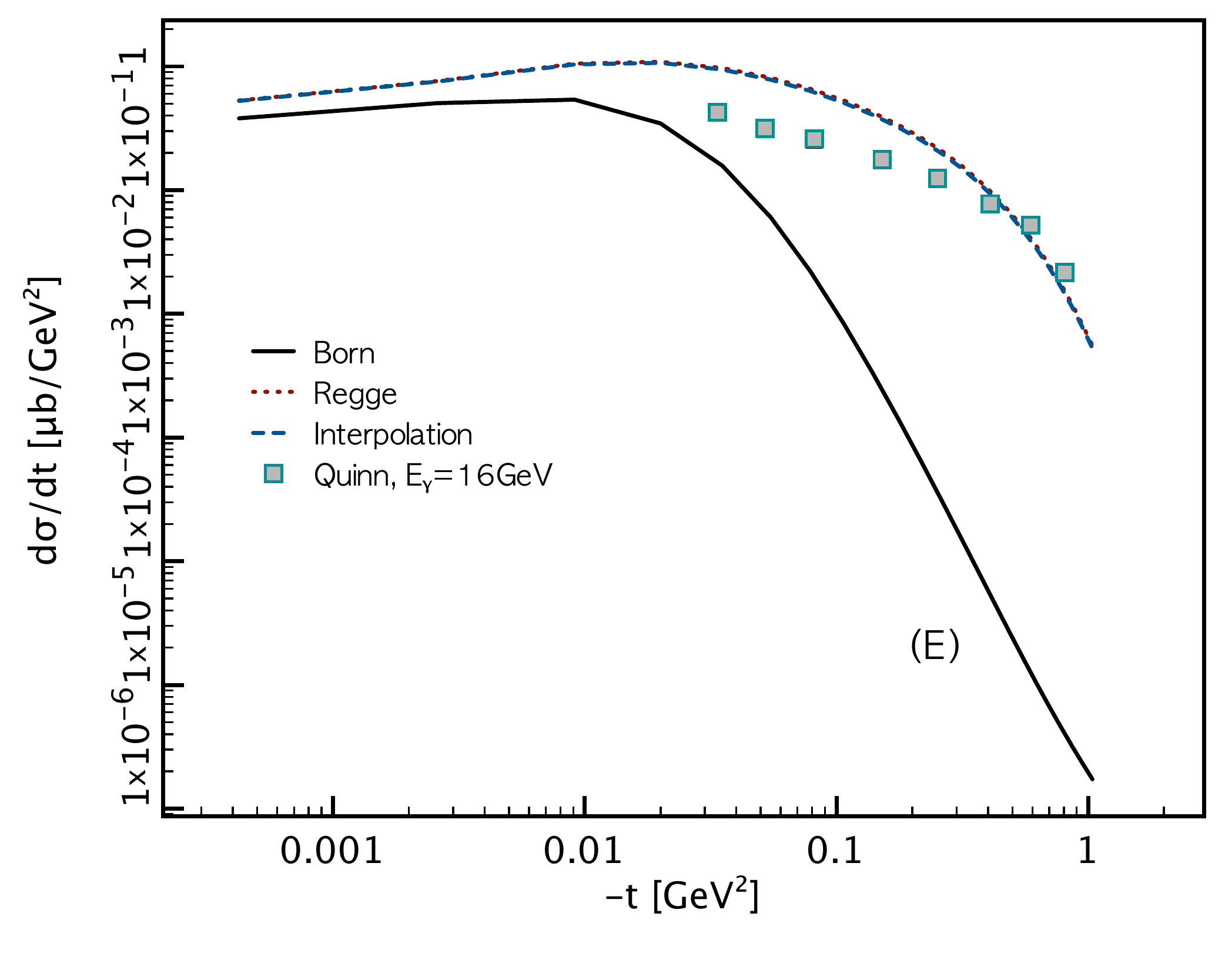}
\caption{(Color online) Momentum-transfer dependence $d\sigma/dt$ [$\mu$b/GeV$^{2}$] for $\gamma p\to\pi^{+}\Delta^{0}$ as a function of $-t$ [GeV$^{2}$] for $E_{\gamma}=16$ GeV. Experimental data are taken from Ref.~\cite{Quinn:1979zp} (Quinn).}
\label{FIG4-1}
\end{figure}

In Fig.~\ref{FIG4}, we present the numerical results of the
momentum-transfer dependence, $d\sigma/dt$ for the higher photon-energy
regions, $E_{\gamma}=5$ (A), $8$ (B), $11$ (C), and $16$ (D) GeV.
The experimental data are taken from
Refs.~\cite{Boyarski:1968dw,Anderson:1976ph,Quinn:1979zp}.
Similarly, we draw the three models, the Born (solid), Regge (dot), and interpolation (dash),
separately. It can be clearly seen that the
interpolation model yields considerably excellent results in comparison
with the experimental data. We also find that the results from the Born
approximation are only reliable below
$-t\approx0.1\,\mathrm{GeV}^{2}$ for all the photon energies.
Beyond this value, we observe that the Regge contribution plays a
critical role as shown in the panels of Fig.~\ref{FIG4}. At the
same time, the interpolation ansatz works very well for these
relatively high energy ranges. For instance, we see that the value
of $\mathcal{R}$ becomes about $0.4$, which indicates that the
contributions from the Regge and Feynman propagators are almost in the
same portion, for $E_{\gamma}=5$ GeV. As expected, much higher
energy such as $E_{\gamma}=16$ GeV, the $\mathcal{R}$ becomes
almost unity, i.e. the Regge propagator prevails almost completely
over the Feynman one. As for the $\pi^+$ photoproduction, we again find a qualitative agreement with the data as shown in the panel E of Fig.~\ref{FIG4-1}, although sizable deviation observed in comparison to that for the $\pi^-$ one. Hereafter, we will show the numerical
results only from the interpolation model, since we have proven that
it has reproduced the experimental
data qualitatively well so far.

\begin{figure}[ht]
\begin{tabular}{cc}
\includegraphics[width=8.5cm]{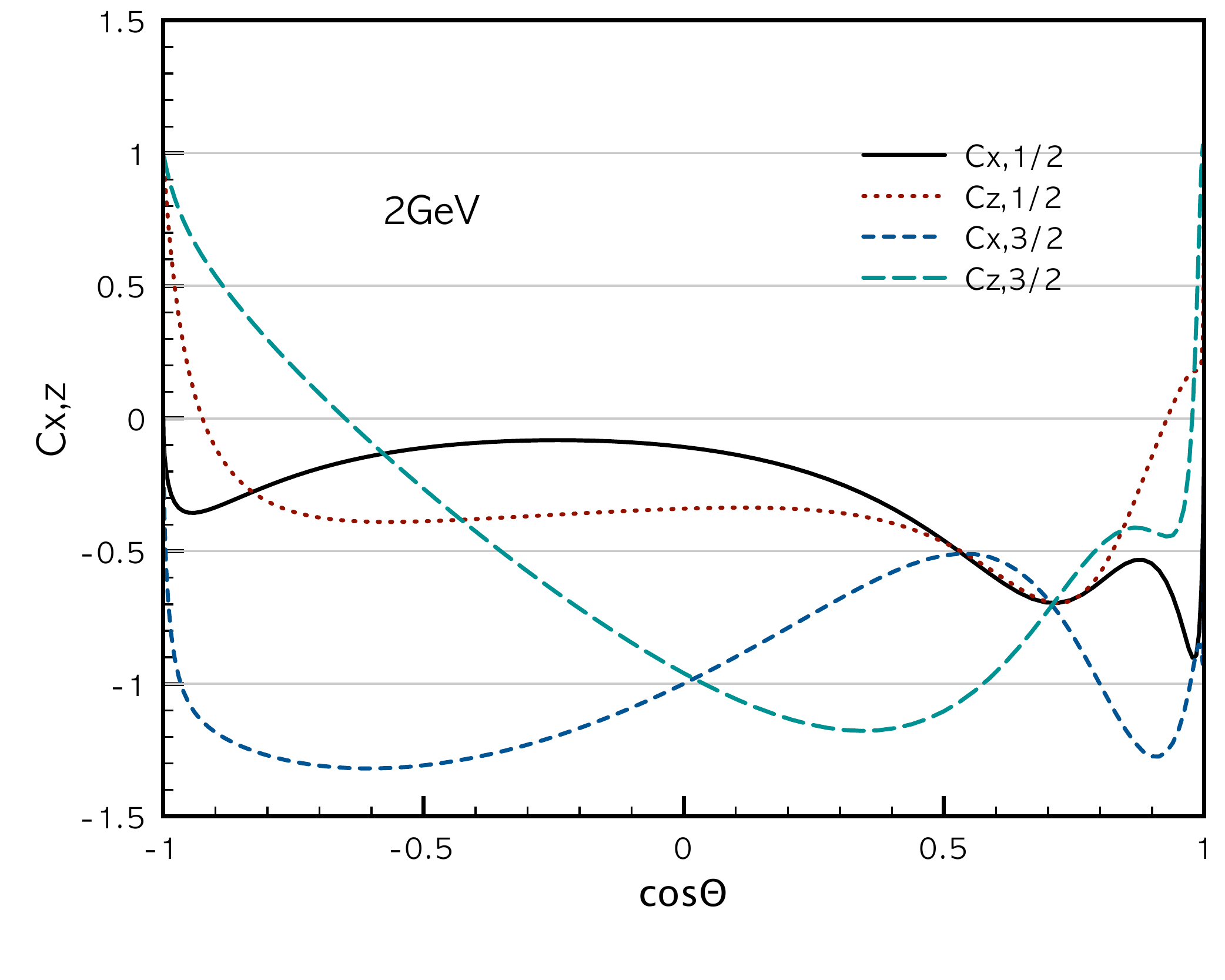}
\includegraphics[width=8.5cm]{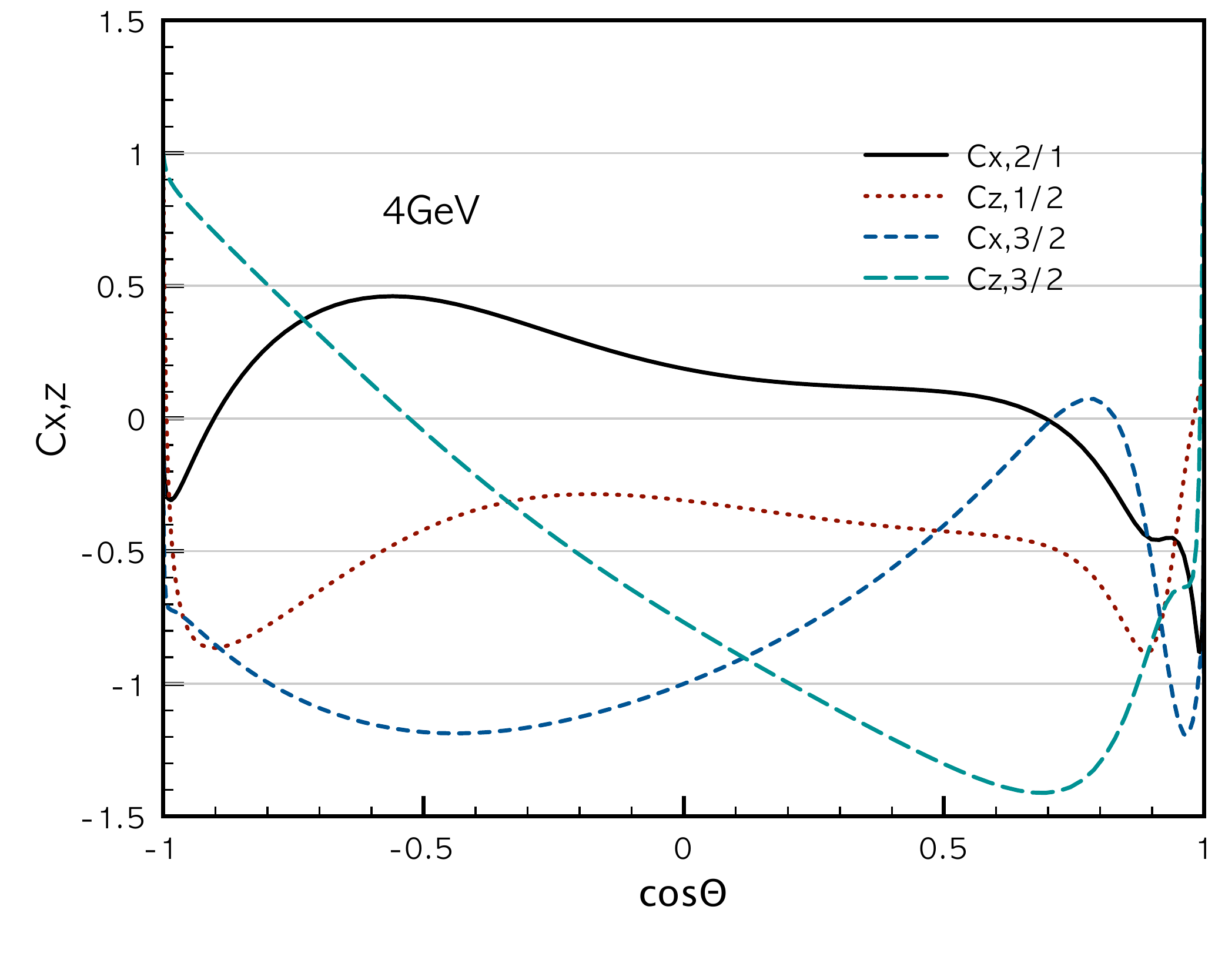}
\end{tabular}
\begin{tabular}{cc}
\includegraphics[width=8.5cm]{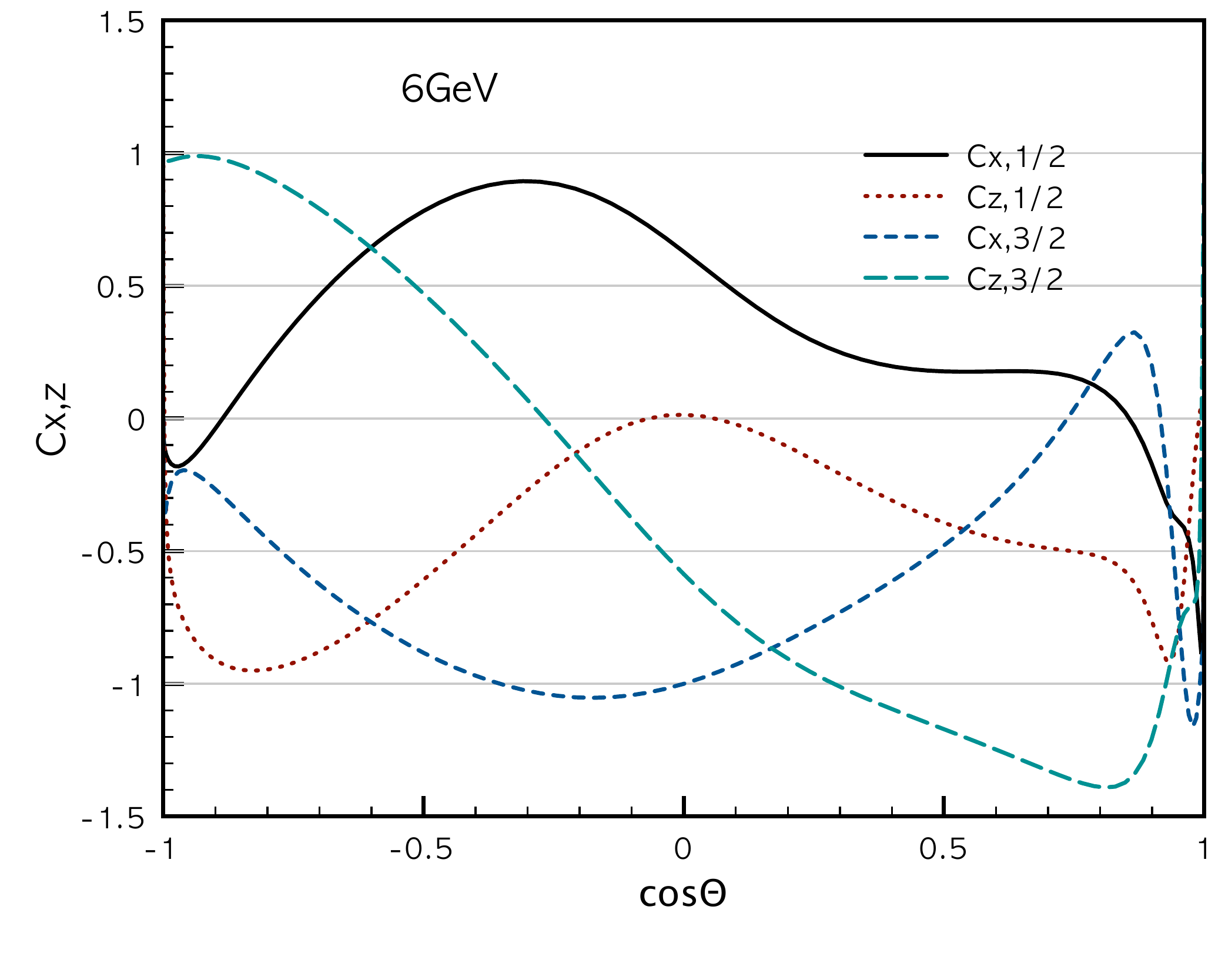}
\includegraphics[width=8.5cm]{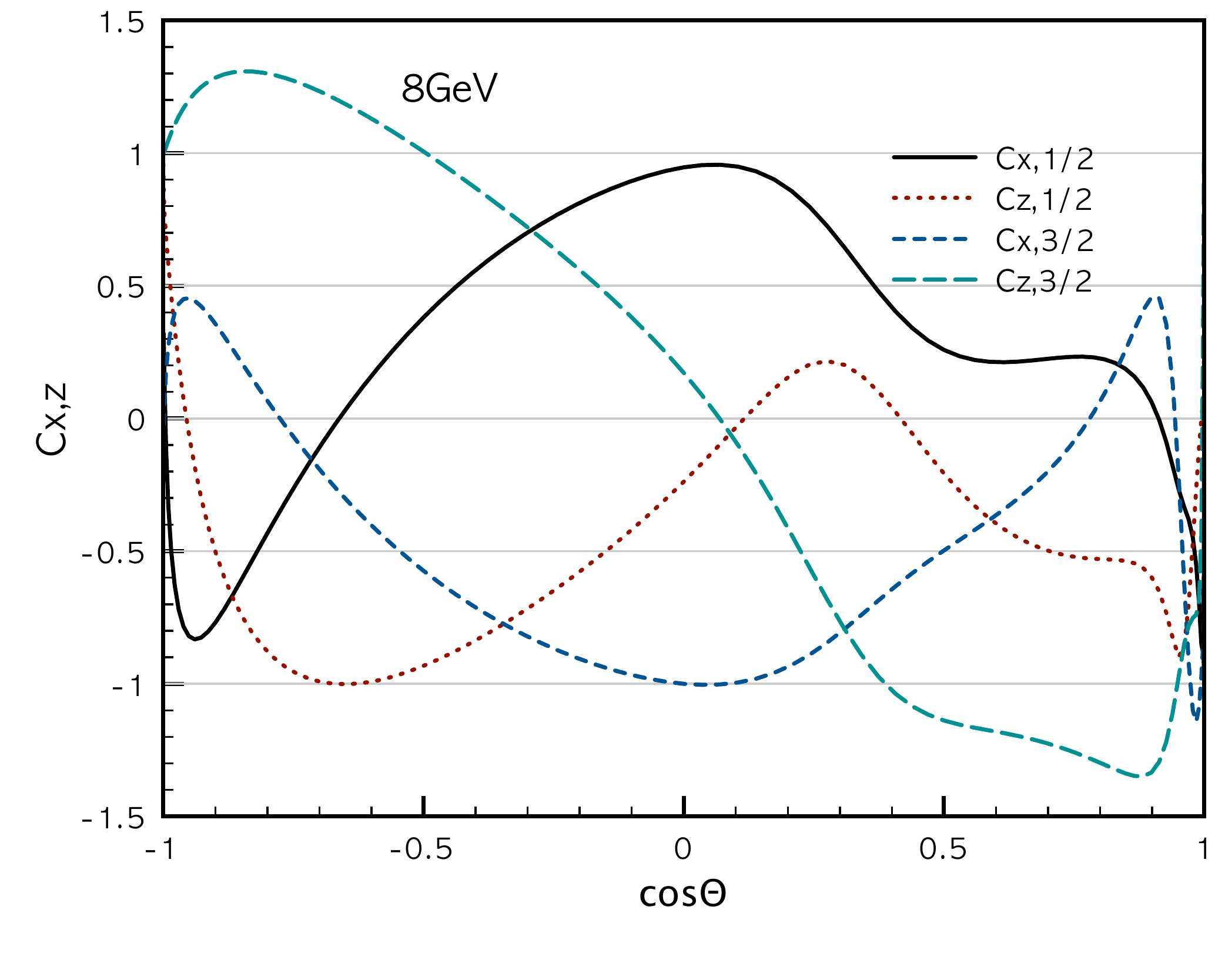}
\end{tabular}
\begin{tabular}{cc}
\includegraphics[width=8.5cm]{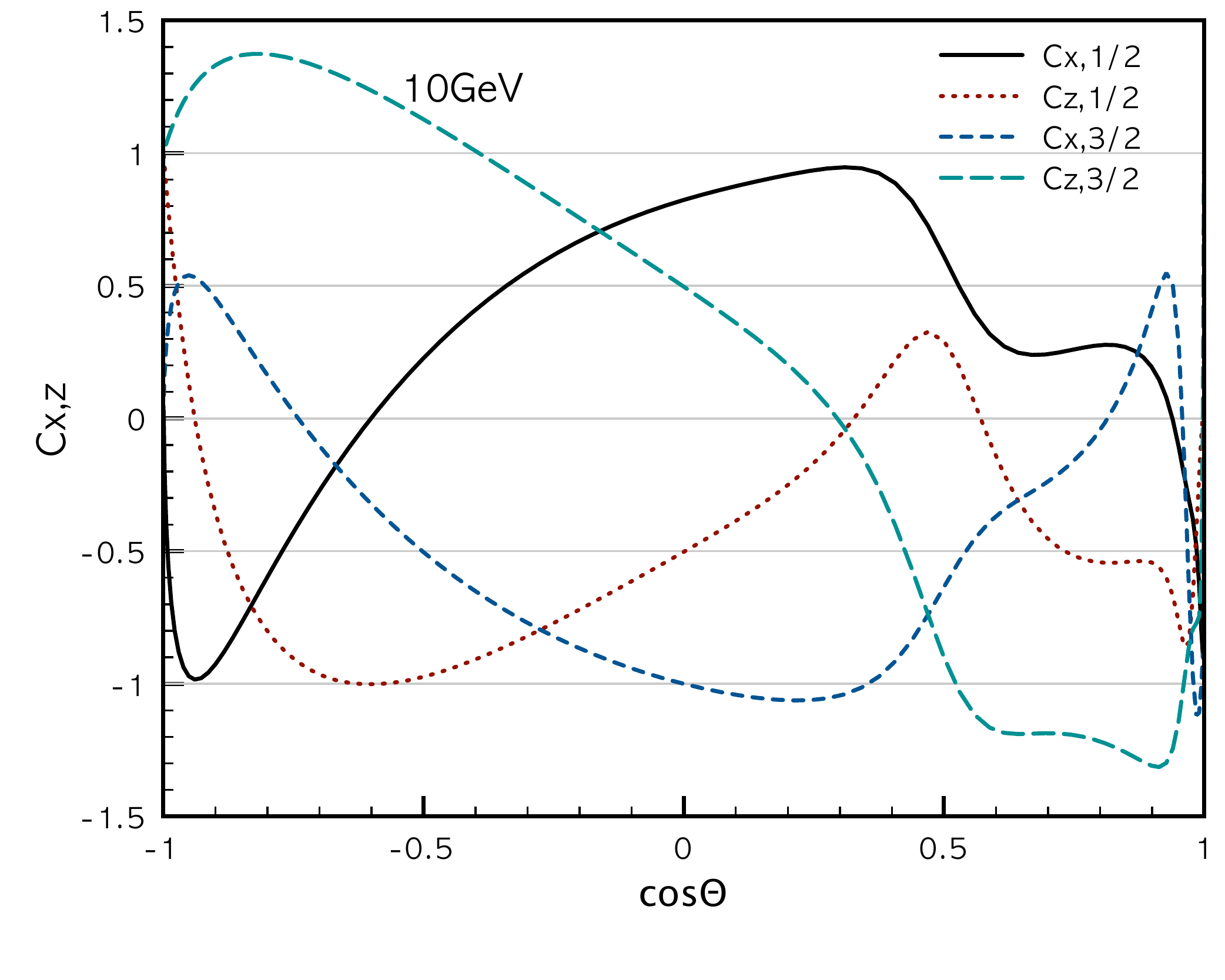}
\includegraphics[width=8.5cm]{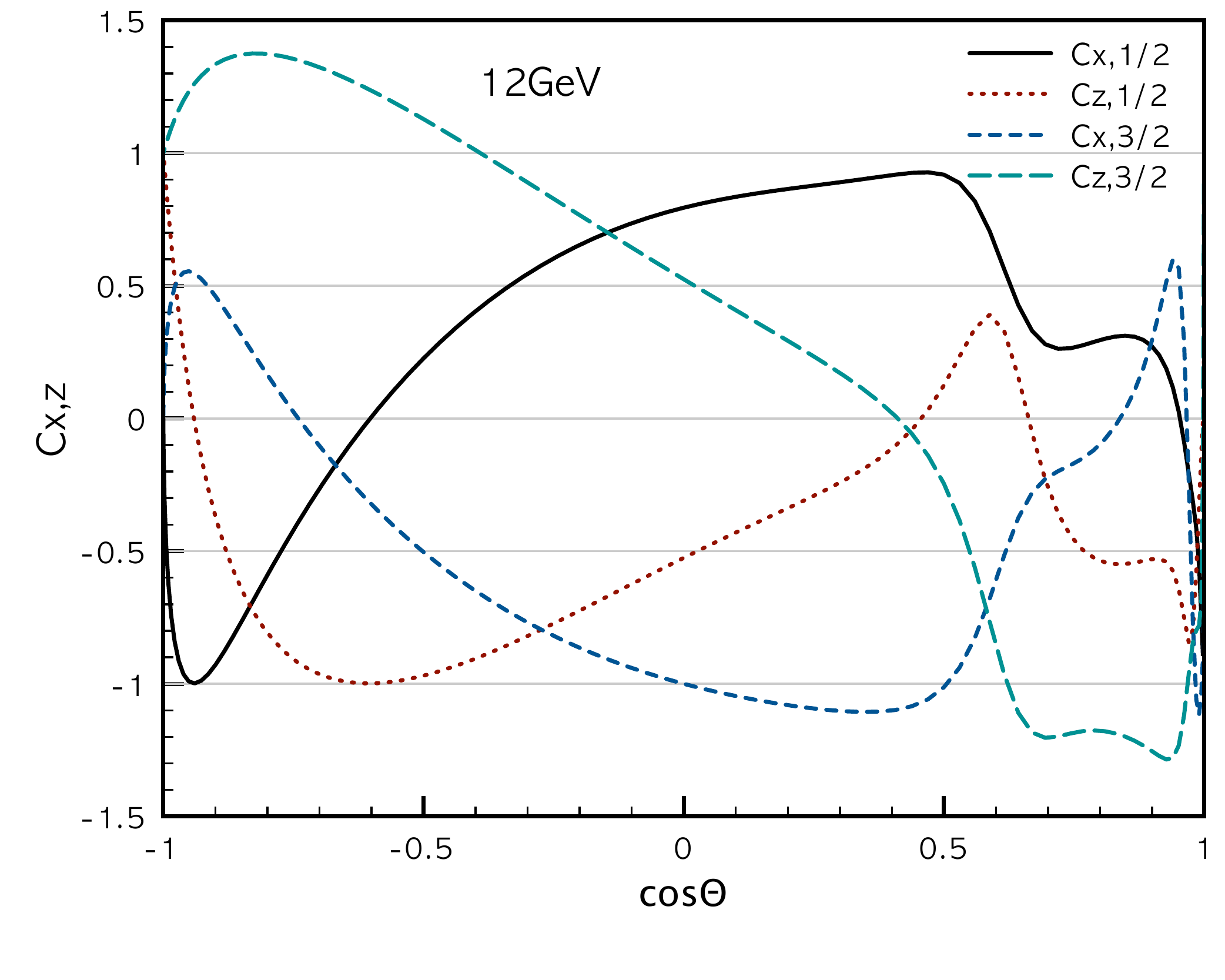}
\end{tabular}
\caption{(Color online) Polarization-transfer coefficients, $C_{x,z}$ for the spin-$1/2$ and $3/2$ components  for $\gamma p\to\pi^{-}\Delta^{++}$ as functions of $\cos\theta$ for different photon energies, $E_{\gamma}=(2,4,6,8,10,12)$ GeV, computed from the interpolation model.}
\label{FIG5}
\end{figure}

Now, we want to discuss the polarization physical quantities especially for the
$\pi^-$ photoproduction, such as the polarization-transfer
coefficients $C_{x,z}$ and related pion-decay distribution. Among the polarization observables in a
meson photoproduction, $C_{x}$ and $C_{z}$ are identified as
the spin asymmetry along the direction of the polarization of the
recoil baryon with the circularly polarized photon beam. First, we
define the polarization transfer coefficients in the
$(x',y',z')$-coordinate, being similar to those for the spin-$1/2$
hyperon-photoproduction as in
Refs.~\cite{McNabb:2003nf,Anisovich:2007bq}:
\begin{equation}
\label{eq:CXZ}
C_{x',|S_{x'}|}=
\frac{\frac{d\sigma}{d\Omega}_{r,0,+S_{x'}}
-\frac{d\sigma}{d\Omega}_{r,0,-S_{x'}}}
{\frac{d\sigma}{d\Omega}_{r,0,+S_{x'}}
+\frac{d\sigma}{d\Omega}_{r,0,-S_{x'}}},\,\,\,\,
C_{z',|S_{z'}|}=
\frac{\frac{d\sigma}{d\Omega}_{r,0,+S_{z'}}
-\frac{d\sigma}{d\Omega}_{r,0,-S_{z'}}}
{\frac{d\sigma}{d\Omega}_{r,0,+S_{z'}}
+\frac{d\sigma}{d\Omega}_{r,0,-S_{z'}}},
\end{equation}
where the subscripts $r$, $0$, and  $\pm S_{x,'z'}$ stand for the
right-handed photon polarization, unpolarized target nucleon, and
polarization of the recoil baryon along the $x'$- or $z'$-axis,
respectively. Since the photon helicity is fixed to be $+1$ here, $C_{x'}$ and $C_{z'}$ measures the polarization transfer to
the recoil baryon. Moreover, $C_{x'}$ and $C_{z'}$ behave as
the components of a three vector so that it can be rotated to the
$(x,y,z)$-coordinate as:
\begin{equation}
\label{eq:ro}
\left(\begin{array}{c}
C_{x}\\C_{z}\end{array}\right)
=\left(
\begin{array}{cc}
\cos{\theta_{K}}&\sin{\theta_{K}}\\
-\sin{\theta_{K}}&\cos{\theta_{K}}
\end{array}
 \right)\left(\begin{array}{c}
C_{x'}\\C_{z'}\end{array}\right),
\end{equation}
where the $(x,y,z,)$-coordinate  stands for that the incident
photon momentum is aligned to the $z$-axis. These polarization
quantities were already investigated  for the spin-$1/2$~\cite{Anisovich:2007bq} and
spin-$3/2$~\cite{Nam:2009cv,Nam:2010au} baryons. The numerical
results for $C_{x,z}$ are presented in Fig.~\ref{FIG5}, employing the interpolation mode onlyl,  for different $E_\gamma$ values. All the results show $C_z=1$ and
$C_x=0$ in the collinear limit, i.e. at $\cos\theta=\pm1$, due to
the helicity conservation. Generally, we observe very complicated
structures for the vicinity $\cos\theta\gtrsim0.5$, since there
happen complicated interferences between the $\pi$-exchange and
contact-term contributions. For $E_\gamma\gtrsim8$ GeV, the shapes
of the curves remain relatively unchanged. This tendency,
negligible changes for the higher photon energies, was already
observed for the $\Lambda(1520)$
photoproduction~\cite{Nam:2009cv,Nam:2010au}. We expect that these
$C_{x,z}$ for the $\Delta$ photoproduction at high energy can be
measured by CLAS collaboration Jefferson laboratory, considering
their upgrading photon-beam energy and high performance of the
circular photon polarization, as already done for the ground state
$\Lambda$ hyperon, i.e. $\Lambda(1192)$~\cite{McNabb:2003nf}.

\begin{figure}[ht]
\includegraphics[width=15cm]{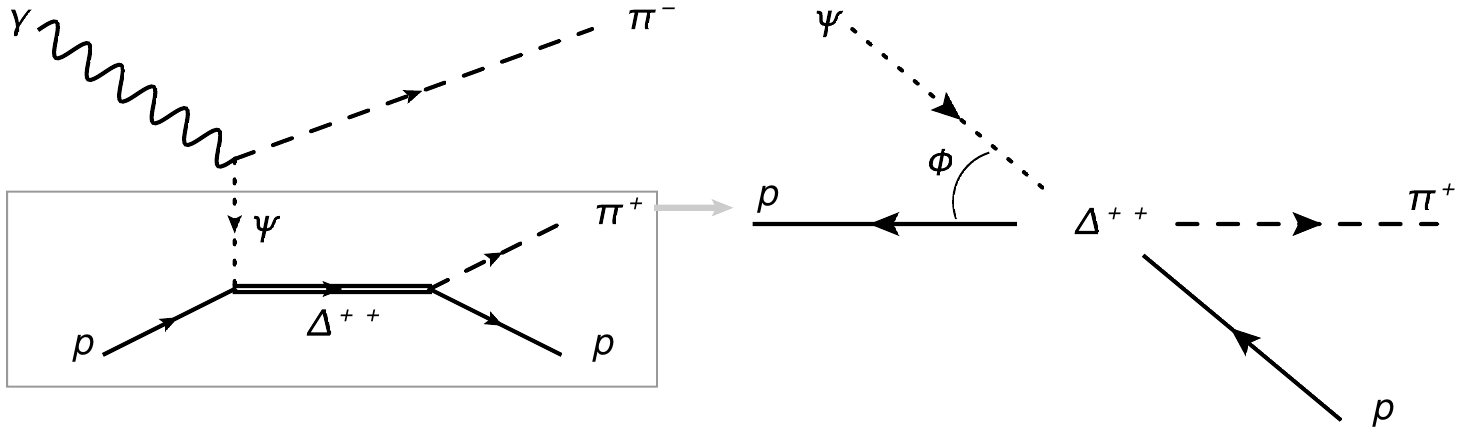}
\caption{Feynman diagram for $\gamma p\to\pi^+\pi^- p$ in the left panel. In the box, the subsequent process, $\psi p\to\Delta^{++}\to\pi p$, is depicted, where $\psi$ indicates a meson exchanged. The diagram in the box can be interpreted using the Gottfried-Jackson frame as shown in the right panel. The angle $\phi$ is defined by that between the initial and final particles.}
\label{FIG6}
\vspace{1cm}
\begin{tabular}{cc}
\includegraphics[width=8.5cm]{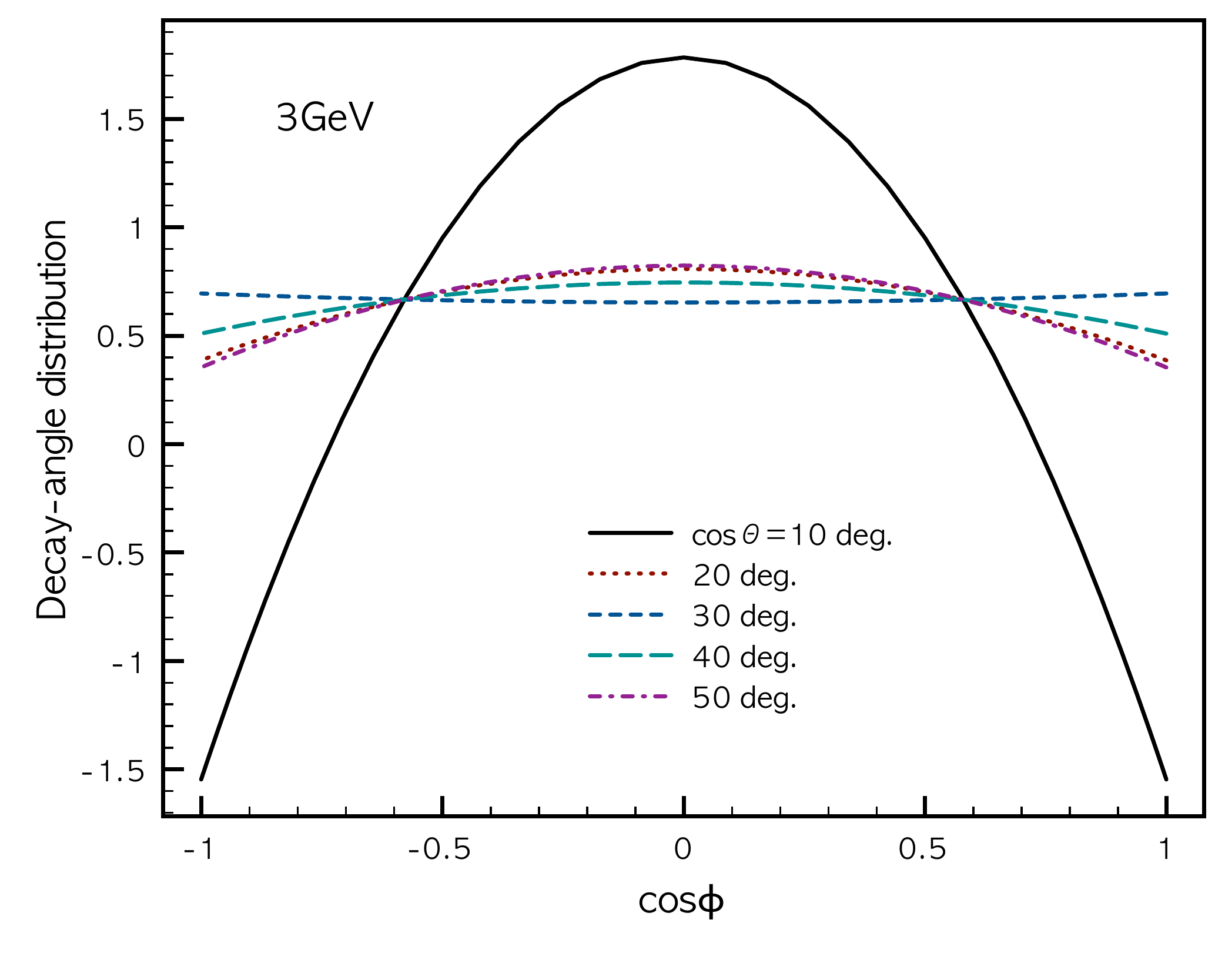}
\includegraphics[width=8.5cm]{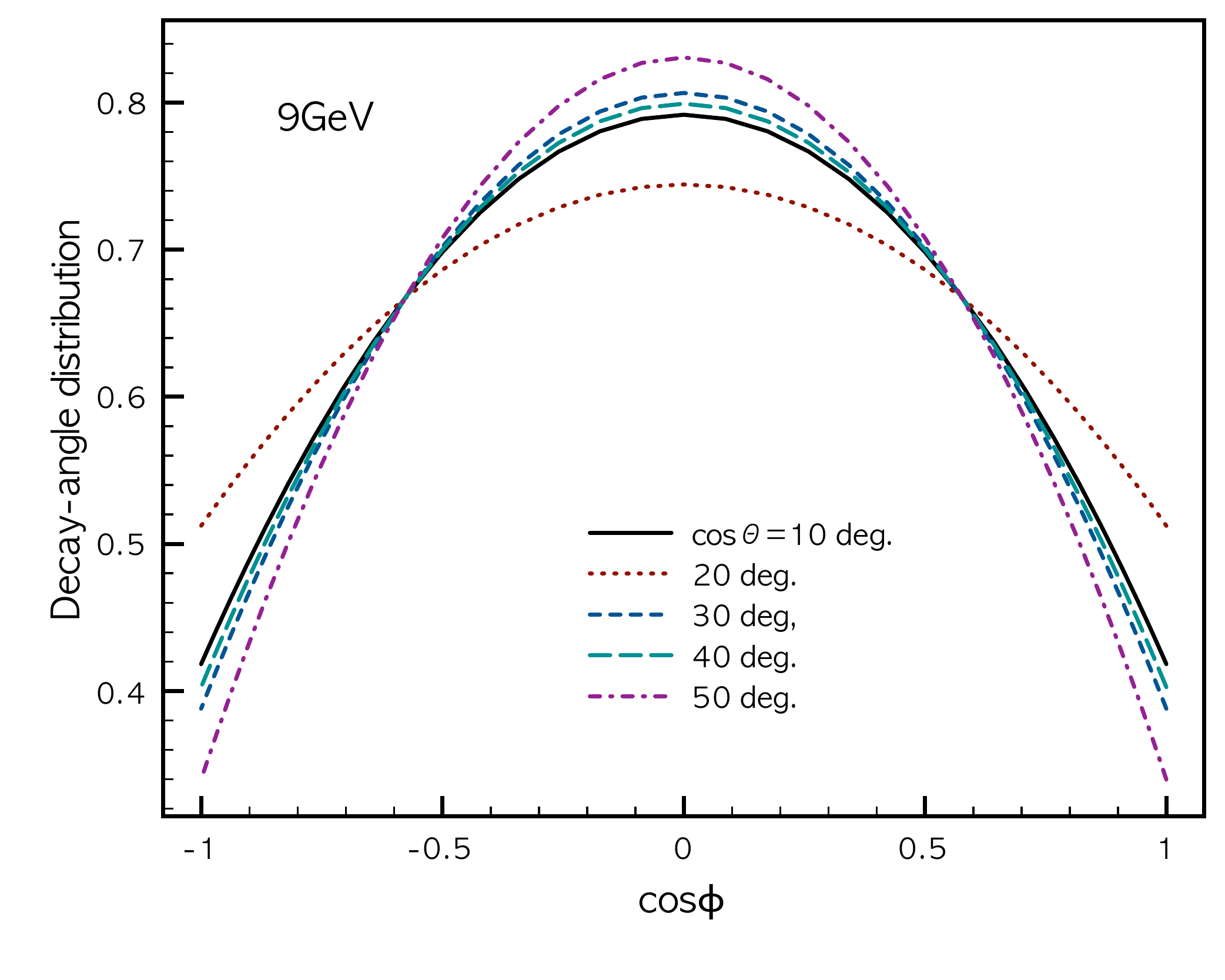}
\end{tabular}
\caption{(Color online) $\pi^{+}$ decay-angle distribution  for $\gamma p\to\pi^{-}\Delta^{++}$ as a function of $\cos\phi$, which is the angle between the $p$ and $\pi^{+}$ decaying from t$\Delta^{++}$ in the Gottfried-Jackson frame (see Fig.~\ref{FIG6}), for two different photon energies $E_{\gamma}=(3,9)$ GeV.}
\label{FIG7}
\end{figure}

Our last topic is the $\pi^{+}$  decay-angle distribution function
for the $\pi^-$ photoproduction process. The decay-angle distribution has
been already discussed experimentally for the $\Lambda(1520)$
photoproduction~\cite{Barber:1980zv,Muramatsu:2009zp}.
Theoretically, it was also explored in the previous
work~\cite{Nam:2009cv,Nam:2010au}. The decay-angle distribution is
the angle distribution of $\pi^{+}$ that decays via
$\Delta^{++}\to\pi^{+}p$ in the $t$-channel helicity frame, i.e.
the Gottfried-Jackson frame~\cite{Schilling:1969um}. Schematic
figures for this frame and kinematics are shown in Fig.~\ref{FIG6}, in which the decay angle $\phi$ is defined. From this distribution
function one can see which meson-exchange is dominating the
production process. According to the spin statistics, the
distribution function becomes proportional to $\sin^{2}\phi$ for
$\Delta^{++}$ in $S_{z}=\pm3/2$, whereas
$\frac{1}{3}+\cos^{2}\phi$ for $\Delta^{++}$ in $S_{z}=\pm1/2$. As
in Ref.~\cite{Muramatsu:2009zp,Barrow:2001ds}, considering all the
possible contributions, we can parametrize the distribution
function as follows:
\begin{equation}
\label{eq:DF}
\mathcal{F}_{\pi^{+}}
=A\sin^{2}\phi+B\left(\frac{1}{3}+\cos^{2}\phi\right),
\end{equation}
where we have used a notation  $\mathcal{F}_{\pi^{+}}$ indicating
the distribution function for convenience. The coefficients $A$
and $B$ stand for the strength of each spin state of $\Delta^{++}$
with the normalization condition $A+B=1$.  In other words, if
$A>B$, one can think that the spin-1 particle exchange or an
equivalent contribution in the $t$ channel dominates the
scattering process, and vice versa for the spin-0 particle
exchange. We note that there can be other hyperon contributions
beside $\Delta^{++}$ so that one can add an additional term to
Eq.~(\ref{eq:DF}) representing the interference effects. However,
we will ignore it for simplicity as done in
Refs.~\cite{Nam:2009cv,Nam:2010au}.

Now, we want to provide theoretical  estimations on
$\mathcal{F}_{\pi^{+}}$. Note that we again only show the numerical results for the interpolation model hereafter. Since the outgoing pion ($\pi^{-}$) carry
no spin, all the photon helicity will be transferred to
$\Delta^{++}$ through the exchanging particle in $t$-channel,
Hence, it is natural to think that the polarization-transfer
coefficients in the $z$ direction should relate to the strength
coefficients $A$ and $B$. Therefore, we can write $A$ and $B$ in
terms of $C_{z,1/2}$ and $C_{z,3/2}$ as follows:
\begin{equation}
\label{eq:AAA}
A=\frac{C_{z,3/2}}{C_{z,1/2}+C_{z,3/2}},\,\,\,\,
B=\frac{C_{z,1/2}}{C_{z,1/2}+C_{z,3/2}},
\end{equation}
which satisfy the normalization condition.  In other words, $A$
denotes the strength that $\Delta^{++}$ is in its $S_{z}=\pm3/2$
state, and $B$ for $S_{z}=\pm1/2$. In Fig.~\ref{FIG7}, we depict
the distribution as a function of $\phi$ for different $\theta=(10\sim50)^\circ$
and $E_\gamma=(3,9)$ GeV. As for the lower energy ($E_\gamma=3$
GeV), the numerical results tell us that the scattering process is
mainly dominated by the spin-$1$ exchange or equivalent
contribution for the very forward scattering angles. Moreover,
this observation indicates that the contact-term contribution
dominates the process, since it can be interpreted equivalently as
a spin-$1$ exchange, and the contact-term one is much more effective
than that of the $\rho$-exchange as discussed above. However, for
the angle around $\theta=30^\circ$, the spin-$0$ exchange, i.e.
$\pi$-meson exchange, contributes significantly, denoted by the
short-dash line in the left panel of Fig.~\ref{FIG7}. Beyond that
angle, again, the contact-term contribution  prevails over that
from the spin-$0$ one. As for the higher energy region, the
situation becomes quite different from the lower energy case. In
the right panel of Fig.~\ref{FIG7}, we show the decay-angle
distribution for $E_\gamma=9$ GeV. We see that, for all the
$\pi^-$ angles, the contact-term contribution dominates the
process, in addition to the tiny $\rho$-meson exchange one, as
understood from the panel D in Fig.~\ref{FIG3}.

\section{Summary and conclusion}

In the present work, we have investigated  the photoproduction of
the charged pion with the $\Delta(1232,3/2^+)$ baryon off the
proton target  at tree level, using the effective Lagrangian
method beyond the resonance region. We took into account the
nucleon- and $\Delta$-pole diagrams corresponding to $s$- and
$u$-channel contributions, respectively. The $\pi$ and $\rho$
exchanges are considered in the $t$-channel. In order for the
gauge invariance of the scattering amplitude, the contact-term
contribution was further included together with the
gauge-invariant scheme for the form factors. The Regge
trajectories for the $\pi$ and $\rho$ mesons are introduced to
describe the high-energy experimental data correctly. We
introduced an ansatz designated to interpolate the Feynman and
Regge propagators. These three different cases were assigned as
the Born, Regge, and interpolation models. It turned out that the
present numerical results are in good agreement with the
experimental data and could provide useful theoretical guides and
estimations for the future experiments. We list the important
observations in the present work as follows:
\begin{itemize}
\item Unpolarized physical quantities such as the energy and
angular dependences of the cross sections are reproduced
qualitatively well beyond the resonance region
$E_\mathrm{cm}\gtrsim2$ GeV for  the interpolation model, in
comparison to the presently available experimental data.
\item We observe a strong peak in  the forward-scattering region
which is suppressed with respect to $E_\gamma$. It also turns out
that this forward peak is generated by the $t$-channel
pseudoscalar-meson ($\pi$) exchange on top of the subleading
contact-term contribution, although the latter gets stronger as
$E_\gamma$ increases. In general, the peak shifts to the very
forward region as $E_\gamma$ increases.
\item The momentum-transfer dependence  of the $d\sigma/dt$ shows
that the present framework is not enough to reproduce the
experimental data for the region $E_\gamma\lesssim3.5$ GeV and
$-t\gtrsim0.05\,\mathrm{GeV}^2$. The observed discrepancy may
indicate the necessity of the further unknown contributions from
the higher-spin meson exchanges in the $t$ channel as well as the
baryon resonances, which are not taken into account in the present
work.
\item On the contrary, the present theoretical framework provides
a very good agreement with the experimental data for the
high-energy region beyond $E_\gamma\approx5$ GeV. Especially, the
ansatz for the Feynman-Regge interpolation plays an important role
to reproduce the data correctly.
\item We also present the  numerical results for one of the
various double polarization observables, i.e., the
polarization-transfer coefficients $C_{x,z}$.  Satisfying  the
condition of the  collinear-limit for $E_\gamma\gtrsim8$ GeV, the observable $C_{x,z}$ remains almost unchanged, whereas
nontrivial structures are shown at the forward scattering regions
due to the interference between the contact-term and $\pi$
exchange contributions.
\item From the numerical results in the $\pi^+$ decay-angle
distribution function, we find that in the low energy region the
spin-$1$ exchange (small $\rho$-meson exchange plus the large
contact-term contribution) and that of the spin-$0$
($\pi$-exchange) competes with each other, depending on the
$\theta$ angles. In contrast, only the spin-$1$ exchanges, such as the contact-term contribution with the small but finite $\rho$-meson exchange give the
dominant contribution to the process in the high-energy region,
i.e. the contact-term contribution. 
\end{itemize}

In conclusion, the present theoretical framework can reproduce the
available experimental data qualitatively well, although 
other resonances and higher-spin meson exchanges are not
considered here. The deviations observed in the region
$E_\gamma\lesssim(3\sim4)$ GeV, therefore, could be improved by
taking into account those contributions. Especially, it also
turns out that the role of the Feynman-Regge interpolation is very
crucial to reproduce the experimental data correctly in the
higher-energy region. Moreover, being different from the
$\Lambda(1520,3/2^-)$ photoproduction  where the contact-term
is dominant, the $t$-channel meson exchange and contact-term contributions  are
competing with each other in the $\Delta(1232, 3/2^+)$, especially
for the lower-energy region. The results from the present
theoretical work would be the useful guides and estimations for
the high-energy photon-beam experiment, such as the future
experiments planned in the LEPS2 at SPring-8 and the CLAS12 at the
Jefferson laboratory, for instance. More sophisticated works, with
other $\pi\Delta$ isospin-channels, nucleon resonances and other
contributions, are under progress and will appear elsewhere.

\section*{Acknowledgment}
The authors are grateful to C.~W.~Kao, A.~Hosaka,  H.-Ch.~Kim for
fruitful discussions. This work was supported by the grant
NRF-2010-0013279 from National Research Foundation (NRF) of Korea.
The numerical calculations were partially performed via SAHO at
RCNP, Osaka University, Japan.  All the figures in the present
work were generated using PLOT~\cite{FIGURE1} and JAXODRAW~\cite{FIGURE3}.

\newpage

\end{document}